\documentclass[12pt,journal,onecolumn]{IEEEtran}

\usepackage{amsmath,amssymb,amsthm,mathtools,bm}
\usepackage{graphicx}
\usepackage{booktabs,array,tabularx,multirow}
\usepackage[table]{xcolor}
\usepackage{cite}
\usepackage{url}
\usepackage[hidelinks]{hyperref}
\hypersetup{
  pdftitle={Inverse Design of Finite-State Channels by Dynamic Measure Matching},
  pdfauthor={Lav R. Varshney},
  pdfsubject={Feedback capacity, constrained systems, inverse Bellman design, and chemical channels},
  pdfkeywords={feedback capacity, finite-state channel, constrained coding, Bellman equation, Q-graph, chemical channel, Meru constant}
}
\usepackage{tikz}
\usepackage{pgfplots}
\pgfplotsset{compat=1.18}
\usepackage{microtype}
\usetikzlibrary{arrows.meta,automata,positioning,calc,fit,backgrounds,shapes.geometric}

\definecolor{IBInk}{gray}{0}
\definecolor{IBShade}{gray}{0.95}
\definecolor{IBGuide}{gray}{0.75}

\tikzset{
  ibarrow/.style={-{Latex[length=2.0mm,width=1.4mm]},line width=0.65pt,draw=IBInk},
  iblegal/.style={ibarrow,solid},
  ibauto/.style={ibarrow,densely dotted,line width=0.8pt},
  ibpremature/.style={ibarrow,dashed,dash pattern=on 3pt off 1.8pt},
  ibbox/.style={draw=IBInk,rounded corners=2.3pt,line width=0.55pt,align=center,
    fill=white,minimum height=7.2mm,minimum width=17mm,font=\scriptsize,inner sep=2.2pt},
  ibstate/.style={circle,draw=IBInk,line width=0.65pt,minimum size=8.5mm,font=\scriptsize,inner sep=0.5pt},
  ibnormal/.style={ibstate,fill=white},
  ibshadow/.style={ibstate,fill=white,dashed,dash pattern=on 2pt off 1.2pt},
  ibmatched/.style={ibstate,fill=white,double,double distance=0.8pt},
  ibnote/.style={draw=IBInk,rounded corners=2pt,fill=white,align=center,font=\scriptsize,inner sep=2.4pt}
}

\newtheorem{theorem}{Theorem}

\newtheorem{proposition}{Proposition}
\newtheorem{corollary}{Corollary}
\newtheorem{lemma}{Lemma}

\newcommand{\head}{\operatorname{head}}
\newcommand{\tail}{\operatorname{tail}}

\newcommand{\Span}{\operatorname{span}}
\newcommand{\Bern}{\operatorname{Bern}}

\title{Inverse Design of Finite-State Channels\\by Dynamic Measure Matching}
\author{Lav R. Varshney}
\begin{document}
\maketitle
\bstctlcite{IEEEcontrol}

\begin{abstract}
Inspired by the optimization approach to biology, we formulate an inverse problem for communication with memory: construct a finite-state channel for which a prescribed signaling behavior supports capacity-achieving codes. Dynamic measure matching is introduced to turn a directed-information converse into conditions on the channel dynamics. These conditions lead to a new notion of dynamic surprise: the log ratio of physical to test output probabilities with a continuation correction for future signaling opportunities. The matching inequalities bound its conditional mean. For fixed architecture, test law, and potential, channel row synthesis under convex physical restrictions is shown to be convex. For every signaling constraint represented by a finite irreducible constrained graph of positive entropy, we construct a channel law under which undesired inputs remain available and behaviorally distinct while incurring a strict information penalty. A noisy refractory family has an exact capacity plateau, followed by an interval in which Shannon and zero-error feedback capacities differ. Explicit zero-error codes attain capacity in the graph construction and on the plateau. In this family, a Gibbs choice maximizes the penalty and makes dynamic surprise equal its conditional mean on every transition, yielding an exact trajectory-wise likelihood identity, a finite-block bound on successful decoding with premature requests, and an exponential strong converse. An exact gap decomposition into action residuals and output-law divergence constrains every reliable capacity-approaching code. Within the strict plateau, positive discrimination exponents for unused dynamics require a positive rate loss, while sparse probing permits consistent estimation at asymptotic capacity. A chemical channel representation illustrates how premature stimulation can alter future responsiveness while leaving the immediate output law unchanged.
\end{abstract}

\begin{IEEEkeywords}
Feedback capacity, finite-state channel, inverse design, measure matching, Bellman equation, directed information, constrained coding, channel identification, $Q$-graph, zero-error capacity.
\end{IEEEkeywords}

\section{Introduction}
\label{sec:intro}

\IEEEPARstart{T}{he} optimization approach to theoretical biology asks which physical mechanisms make an observed behavior optimal \cite{Alexander1996,ParkerMaynardSmith1990,VarshneyKusumaGoyal2016}, as a way to generate hypotheses that can be experimentally tested. We study this inverse problem for communication with memory: construct a finite-state channel for which a prescribed causal signaling behavior supports capacity-achieving codes. More broadly, this formulation is related to inverse optimization, which takes decisions as input and seeks model parameters under which they are optimal or nearly optimal \cite{ChanMahmoodZhu2025}.

In biological information theory, a resource law and a noise model together determine which distributions of physical states use available resources most effectively \cite{LevyBaxter1996}. As an example, maximizing information storage under a volume constraint has been used to account for noisy synapses, broad synaptic-weight distributions, and sparse connectivity \cite{VarshneyThesis2006,VarshneySjostromChklovskii2006,VarshneyChklovskii2007}. Related optimization-based explanations occur outside biology. Optimal channel network models relate river drainage organization to energy dissipation and dynamically accessible local optima \cite{RinaldoEtAl2014}. In inverse statistical mechanics, interactions are chosen to favor a prescribed structure: Rechtsman, Stillinger, and Torquato designed potentials for targeted self-assembly \cite{RechtsmanStillingerTorquato2005}, and Cohn and Kumar proved ground-state optimality for selected configurations under constructed potentials \cite{CohnKumar2009}.

In our inverse problem, the target behavior is fixed before the channel is designed. Within an admissible physical architecture, we choose a channel kernel $W$ and seek two operational conclusions: codes implementing the prescribed behavior attain a specified rate, and every reliable sequence of feedback codes on the resulting channel has asymptotic rate at most that value. To avoid triviality, inputs omitted by the target behavior remain physically available. A stronger design objective is to make specified departures incur a positive information penalty, so that every reliable code approaching capacity uses them with vanishing frequency.

\begin{figure*}[t]
\centering
\begin{tikzpicture}[font=\small]
\node[ibbox,fill=white,text width=29mm,minimum height=13mm] (target) at (0,0)
 {Prescribed process\\and signaling code};
\node[ibbox,fill=white,text width=29mm,minimum height=13mm] (match) at (4.0,0)
 {Test process $T$\\and potential $v$};
\node[ibbox,fill=white,text width=29mm,minimum height=13mm] (rows) at (8.0,0)
 {Channel rows $W$\\and residuals $\Delta$};
\node[ibbox,fill=white,text width=29mm,minimum height=13mm] (cap) at (12.0,0)
 {Exact capacity\\and stability};
\node[ibbox,fill=white,dashed,text width=29mm,minimum height=10mm] (arch) at (8.0,-1.55)
 {Physical architecture\\supports and costs};
\draw[ibarrow] (target)--(match);
\draw[ibarrow] (match)--(rows);
\draw[ibarrow] (rows)--(cap);
\draw[ibarrow] (arch)--(rows);
\draw[iblegal] (target.south) -- (0,-2.55) --
 node[pos=.5,below=2pt,font=\scriptsize] {separate coding lower bound} (12,-2.55) -- (cap.south);
\node[font=\scriptsize,IBInk] at (6.0,.94) {inverse design from all-policy Bellman inequalities};
\end{tikzpicture}
\caption{Inverse channel design by dynamic measure matching. A prescribed signaling process and coding construction provide an achievable rate. A test law and continuation potential give conditions for selecting channel rows within the allowed physical architecture. The Bellman inequalities bound every causal strategy, while the prescribed code attains the bound. Strict residuals quantify the information-rate loss caused by departures from the prescribed behavior.}
\label{fig:framework}
\end{figure*}
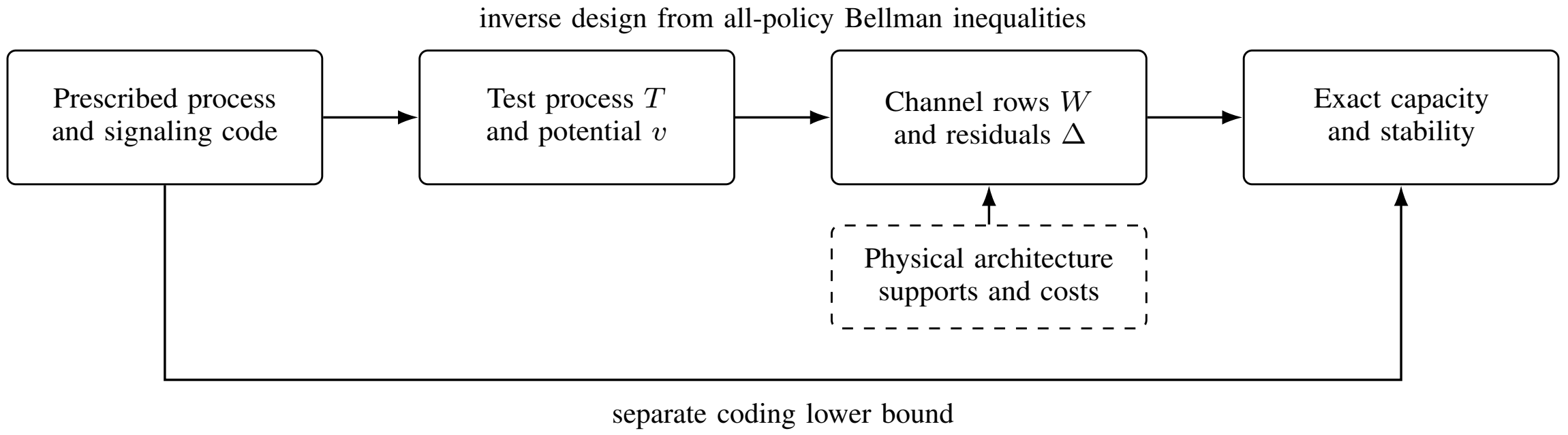

Memory couples the present observation law to the state from which future signals will be sent. We use a test output process $T$ and a continuation potential $v$ to express this dependence through \emph{dynamic surprise}: the log ratio of physical to test output probabilities plus the change in $v$. Two actions with the same immediate output law can differ through their effects on future signaling opportunities. The resulting Bellman inequalities must hold for every receiver context compatible with a physical state, including contexts reached after departures from the desired behavior. Their residuals $\Delta$ quantify the information opportunity lost through an action. Figure~\ref{fig:framework} shows how these inequalities and a separate coding construction enter the inverse problem.

The temporal effect addressed here has a biochemical interpretation. A premature stimulus may be indistinguishable from an ordinary recovery stimulus at the present use, yet reduce subsequent responsiveness. Thomas and Eckford derive the BIND channel from ligand binding and unbinding dynamics and show that feedback does not increase its capacity \cite{ThomasEckford2016}. Eckford and Thomas subsequently model intensity-driven signal-transduction receptors as input-driven finite-state Markov chains and characterize their information rates and capacities \cite{EckfordThomas2018}. Experimentally, nanomolar agonist can inhibit later P2X3 receptor responses through rebinding to desensitized receptors before recovery \cite{PrattEtAl2005}. As we will see in the sequel, such behavior motivates the normal and shadow recovery states of our key refractory family construction. These are clocked finite-state abstractions; a kinetic interpretation requires independently measured conformations, recovery-time distributions, input dependence, and observation noise. Causal output feedback corresponds to a controller that observes the response before selecting the next stimulus.

\subsection{Main results}

The design principle follows from the graph-structured directed-information duality bound of Huleihel, Sabag, Aharoni, and Permuter \cite{HuleihelSabagAharoniPermuter2025}. For physical state $s$, receiver context $q$, and input $x$, let $W_Y(\cdot\mid s,x)$ be the conditional output law and $T(\cdot\mid q)$ the test law. The context is recursively generated from past outputs. With $D$ denoting relative entropy in bits, define
\begin{equation}
 r_T(s,q,x)=D\bigl(W_Y(\cdot\mid s,x)\|T(\cdot\mid q)\bigr).
 \label{eq:intro-reward}
\end{equation}
Let $S^+$ and $Q^+$ be the next physical state and receiver context. A potential $v(s,q)$ gives the dynamic matching condition
\begin{equation}
 r_T(s,q,x)+\mathbb E v(S^+,Q^+)-v(s,q)\le\rho,
 \label{eq:intro-match}
\end{equation}
where the expectation is conditional on $(s,q,x)$ and $\rho$ is a candidate capacity bound in bits per channel use. The left side is the conditional mean of dynamic surprise, developed in Section~\ref{sec:matching}. This condition must hold at every pair reachable under every physical input. Its residual $\Delta((s,q),x)$ is $\rho$ minus the left side. Desired actions attain equality, while specified undesired actions may have a strictly positive residual.

The residual enters an exact decomposition of the gap between directed information and the Bellman bound. A channel is unifilar when its next state is determined by its current state, input, and output. For such a channel, let $X_t,Y_t$ be the input and output at time $t$, let $X^n=(X_1,\ldots,X_n)$ and $Y^n=(Y_1,\ldots,Y_n)$, and write $I(X^n\!\to Y^n)$ for directed information. Let $Z_t=(S_t,Q_t)$ collect the physical state and context after use $t$, with fixed known initial pair $z_0=(s_0,q_0)$. For the actual and test output laws $P_{Y^n}$ and $T_{Y^n}$,
\begin{align}
 &n\rho+v(z_0)-\mathbb E v(Z_n)-I(X^n\!\to Y^n)\nonumber\\
 &\quad=\mathbb E\sum_{t=1}^n\Delta(Z_{t-1},X_t)
       +D(P_{Y^n}\|T_{Y^n}).
 \label{eq:intro-gap}
\end{align}
Thus the same inequalities both bound all causal strategies and account for their loss relative to the target rate. When an explicit desired code attains $\rho$, they establish capacity. With the state-update topology, test law, and potential fixed, selecting channel rows to satisfy these inequalities is a convex problem under convex physical restrictions. When only output supports restrict a row, a minimax formula gives it as a geometric mixture of test laws tilted by continuation values, subject to compatibility with every receiver context. The construction chooses a single physical row for all those contexts.

We give two constructive realizations: one for constrained graphs and one for noisy refractory dynamics. For every finite irreducible constrained graph of positive entropy, its Perron--Frobenius eigenvector and maximum-entropy path measure determine a soft finite-state channel whose capacity equals the constrained entropy. A violating input initiates an autonomous path; its Bellman penalty equals the entropy of the path over which control has been surrendered.

The more compact noisy construction uses normal and shadow recovery states. Let $a\in[0,1]$ be the probability of an additional response after a premature request. The prescribed refractory rate equals Shannon feedback capacity exactly for $0\le a\le a_{\rm c}$, where $a_{\rm c}\in(0,1)$ is the nonzero root of $H_2(a)=a$ and $H_2$ denotes binary entropy (Theorem~\ref{thm:capacityplateau}). Beyond this plateau, occasional premature requests improve the achievable rate, while zero-error feedback capacity remains at the prescribed rate for $a<1$. The two feedback capacities therefore differ on $a_{\rm c}<a<1$. Explicit zero-error codes attain capacity in the graph realization and throughout the refractory plateau.

The refractory construction also identifies a Gibbs choice that maximizes the penalty for premature requests and makes dynamic surprise equal its conditional mean on every possible transition. The resulting trajectory-wise likelihood identity gives a finite-block bound on the joint probability of correct decoding and at least a specified number of premature requests (Theorem~\ref{thm:trajectorymatching}). It yields an exponential strong converse: correct-decoding probability decreases exponentially at any fixed rate above capacity. At rates tending to capacity, successfully decoded trajectories with a fixed positive premature-request density are exponentially rare.

Strict matching also determines what near-capacity behavior can reveal about the mechanism. Equation~\eqref{eq:intro-gap} forces both the average action residual and normalized output-law divergence to vanish along reliable capacity-approaching codes. In the refractory family, the prescribed behavior leaves the parameter-dependent channel row unvisited. Adaptive experiments acquire parameter information through premature requests, whose positive residual links discrimination to communication deficit. Within the strict capacity plateau, a positive discrimination error exponent requires a positive rate loss; a growing but sublinear number of probes permits consistent estimation at asymptotic capacity. These are consequences of the mechanism selected by inverse design, and they distinguish capacity attainment from identification of that mechanism.

\subsection{Prior work}

The inverse viewpoint has direct precedents. Gastpar characterizes probabilistic matching for memoryless source--channel communication \cite{GastparRimoldiVetterli2003,GastparThesis2002}, which we interpret from an inverse design perspective \cite{VarshneyThesis2006,VarshneySjostromChklovskii2006,VarshneyChklovskii2007}. Gorantla and Coleman study inverse-optimal causal coding, including an interpretation of the trapdoor mechanism as an inverse-control-optimal coordination strategy \cite{GorantlaColeman2011}. Ghatrani and Ghate study inverse Markov decision problems in which transition probabilities are chosen to make a prescribed policy optimal \cite{GhatraniGhate2023}. In the present channel problem, the information reward changes with the kernel being designed, and one physical row must satisfy the inequalities for every compatible receiver context. Capacity equality also requires an operational coding construction.

Directed information and dynamic programming provide the forward framework for channels with feedback \cite{Massey1990,Kramer1998,TatikondaMitter2009,PermuterWeissmanGoldsmith2009}. The trapdoor channel is a classical exact example \cite{PermuterCuffVanRoyWeissman2008}; the generalized chemical channel was studied further in \cite{WuAnastasopoulos2016}. Sabag, Permuter, and Pfister introduced output-recursive $Q$-contexts and a single-letter bound for unifilar channels \cite{SabagPermuterPfister2017}. Subsequent work developed graph-based encoders \cite{SabagHuleihelPermuter2020} and computable dual bounds \cite{HuleihelSabagEtAl2021}. The directed-information duality bound \cite{HuleihelSabagAharoniPermuter2025} is the immediate technical basis of our design conditions; \cite{TsurSabagEtAl2026} surveys this program. The relation between positive eigenvectors and entropy- or relative-entropy-based control is established \cite{Parry1964,Todorov2006} and provides a constructive input to the graph realizations. The inverse problem has not been addressed in this literature.

The statistics of good codes provide another antecedent. Han and Verd\'u studied approximation of output statistics \cite{HanVerdu1993}; Shamai and Verd\'u established convergence properties of empirical distributions of capacity-approaching codes, including a class of channels with memory \cite{ShamaiVerdu1997}; and Polyanskiy and Verd\'u obtained quantitative output-distribution bounds for memoryless channels at nonvanishing error probability \cite{PolyanskiyVerdu2014}. Our exact identity retains a local action-residual term together with output-process divergence for arbitrary causal strategies. These terms are then used to design channel dynamics and constrain near-capacity behavior.

Mayya, Mainsah, and Reeves obtained the $(d,\infty)$ constrained entropy for a noiseless refractory model of the P300 speller, including recovery-resetting inputs \cite{MayyaMainsahReeves2016}. Input-constrained feedback capacities have also been obtained for erasure and binary-input binary-output channels \cite{SabagPermuterKashyap2016,SabagPermuterKashyap2018,RameshwarKashyap2022}. Our refractory example adds a noisy trapdoor-type readout, an inverse-designed off-policy mechanism, and a uniform residual over every physical strategy. The range over which the prescribed behavior remains optimal is new. The second and fourth Meru constants associated with Pi\.{n}gala's \emph{Meru-prast\={a}ra} \cite{Varshney2008Meru} give two explicit instances.

At the end, we note that optimal signaling and mechanism identification raise distinct questions. Tavakoli, Nguyen, and Bose use mutual-information optimality conditions to estimate a memoryless channel parameter and its capacity-achieving input law from output observations \cite{TavakoliNguyenBose2026}. Wu and Joudeh characterize communication-rate and channel-discrimination exponent tradeoffs in a memoryless joint communication and sensing model \cite{WuJoudeh2024}. Within the strict capacity plateau of our refractory family, observations of the unknown parameter occur only in states reached through actions with a positive Bellman residual. This yields a converse linking discrimination to communication deficit.

\subsection{Organization}

The remainder of the paper is organized as follows. Section~\ref{sec:model} specifies the inverse problem and its operational requirements. Section~\ref{sec:matching} develops dynamic matching, the exact rate-gap identity, and its consequences for capacity-approaching codes. Section~\ref{sec:synthesis} gives convex row synthesis. Section~\ref{sec:constraints} constructs channels from constrained systems. Section~\ref{sec:refractory} develops the refractory realization, its exact capacity plateau, and the trajectory-wise converse at the Gibbs choice. Section~\ref{sec:interpretation} derives the communication and identification consequences of its off-policy dynamics.

\section{Problem formulation}
\label{sec:model}

All alphabets are finite and all logarithms are base two unless explicitly written as $\ln$. For distributions $P,Q$, $D(P\|Q)$ is relative entropy with its usual support convention. We write $H_2$ for binary entropy, $\Bern(a)$ for the binary law assigning probability $a$ to one, $D_{\rm B}(a\|b)=D(\Bern(a)\|\Bern(b))$, and $\Span(v)=\max v-\min v$. Total-variation distance is $\|P-Q\|_{\rm TV}=\tfrac12\sum_y|P(y)-Q(y)|$. We write $x_i^j=(x_i,\ldots,x_j)$ for an interval of a sequence. Initial channel and receiver-context states are fixed and known.

Let $\mathcal S$, $\mathcal X$, and $\mathcal Y$ be the physical state, input, and output alphabets. A unifilar finite-state channel is specified by
\begin{equation}
 W_Y(y\mid s,x),\quad s^+=f(s,x,y).
 \label{eq:kernel}
\end{equation}
Here $f:\mathcal S\times\mathcal X\times\mathcal Y\to\mathcal S$ is the deterministic channel-state update: after input $x$ in state $s$ and observation $y$, the next state is $s^+=f(s,x,y)$. Equivalently, $S_t=f(S_{t-1},X_t,Y_t)$. The state $S_{t-1}$ is therefore known to the encoder from $(s_0,X^{t-1},Y^{t-1})$. A length-$n$ feedback code has a uniform message $M\in\{1,\ldots,M_n\}$, causal inputs $X_t=\varphi_t(M,Y^{t-1})$, and an estimate $\widehat M(Y^n)$. Write $P_e^{(n)}=\Pr\{\widehat M\ne M\}$ for the decoding-error probability. For a randomized encoder, write $X_t=\varphi_t(M,K,Y^{t-1})$, where $K$ is a private random seed independent of the message and channel noise. The feedback capacity $C_{\mathrm{fb}}$ and zero-error feedback capacity $C_{0,\mathrm{fb}}$ are the suprema of rates achievable with vanishing and zero error, respectively, using codes with fixed blocklengths \cite{Shannon1956}. A cost-constrained code has a specified nonnegative physical cost $b(s,x)$ and satisfies $n^{-1}\mathbb E\sum_{t=1}^n b(S_{t-1},X_t)\le B$, where $B$ is the budget. The corresponding capacities are denoted $C_{\mathrm{fb}}(B)$ and $C_{0,\mathrm{fb}}(B)$. All achievable constructions below can be chosen deterministic once a message is fixed.

An architecture specifies $(\mathcal S,\mathcal X,\mathcal Y,f,s_0)$ and allowed output supports $\mathcal A_{s,x}$. Additional physical restrictions determine a class $\mathcal W_{\mathrm{phys}}$ of admissible channel kernels. The prescribed behavior may be a finite-state causal policy or a class of graph paths with a causal implementation and decoder. Given a target rate $C_\star$, the inverse problem is to find $W\in\mathcal W_{\mathrm{phys}}$ such that codes implementing that behavior attain $C_\star$ and arbitrary reliable feedback codes on the resulting channel have asymptotic rate at most $C_\star$. If a cost budget is specified, both statements use that budget. The desired codes give an operational lower bound, and a converse must cover all physical inputs, including those unused by the target behavior. A realization therefore establishes $C_{\mathrm{fb}}=C_\star$, or $C_{\mathrm{fb}}(B)=C_\star$ in the cost-constrained case. When the desired codes have zero error, the corresponding zero-error capacity has the same value.

Capacity attainment allows other behaviors to attain the same rate. A stronger design objective is to assign a uniform positive Bellman residual to specified undesired state--action pairs. Under the conditions of Section~\ref{sec:matching}, every reliable capacity-approaching code then visits those pairs with vanishing expected frequency. We call a realization \emph{soft} when an input omitted by the desired behavior remains available and, under a common continuation, induces a future output law different from that induced by a desired input at the same state. Inputs in an autonomous mode may be equivalent; the distinction concerns the initiating action. In the general construction, a ``violation'' is a locally undesired physical action that remains an admissible channel input.

We consider both channel-row synthesis within a fixed architecture and constructions that also select the architecture. In the row-synthesis problem of Section~\ref{sec:synthesis}, the architecture, output supports, receiver-context update, test law, and potential are fixed. The design variables are the physical output rows, and closed convex restrictions on those rows preserve convexity. The joint choice of architecture, test process, and potential depends on the target behavior and the physical restrictions. In Section~\ref{sec:constraints}, a finite irreducible constrained graph of positive entropy is the target, and we construct an architecture with autonomous modes together with a test law and potential. The graph construction is allowed to add autonomous states.

A finite $Q$-graph has a context alphabet $\mathcal Q$ and deterministic update $g:\mathcal Q\times\mathcal Y\to\mathcal Q$. Starting from $q_0$, its context and test output law are
\begin{equation}
 Q_t=g(Q_{t-1},Y_t),\quad
 T_{Y^n}(y^n)=\prod_{t=1}^nT(y_t\mid q_{t-1}),
 \label{eq:test}
\end{equation}
where $q_t=g(q_{t-1},y_t)$ along the displayed output string. The pair $Z_t=(S_t,Q_t)$ is used in the converse. Its all-policy reachable set is the smallest set $\mathcal Z_{\mathrm{all}}$ containing $(s_0,q_0)$ and satisfying
\begin{equation}
 \begin{aligned}
 &(s,q)\in\mathcal Z_{\mathrm{all}},\quad W_Y(y\mid s,x)>0\\
 &\quad\Longrightarrow\ (f(s,x,y),g(q,y))\in\mathcal Z_{\mathrm{all}}.
 \end{aligned}
 \label{eq:closure}
\end{equation}
for every $x$. This finite closure includes states reached after departures from the target behavior. Every test law used below satisfies
\begin{equation}
 W_Y(\cdot\mid s,x)\ll T(\cdot\mid q)
 \quad\text{for all }(s,q)\in\mathcal Z_{\mathrm{all}},\ x\in\mathcal X.
 \label{eq:domination}
\end{equation}
Zeros in $T$ are permitted only when the corresponding output is impossible from every compatible physical row. A receiver context is determined by observations and is not a physical control variable. Consequently, a designed row $W_Y(\cdot\mid s,x)$ must meet the matching inequalities for every context compatible with $s$.

For an induced causal input law, let the directed information be
\begin{equation}
 I_n=I(X^n\!\to Y^n)=\sum_{t=1}^n I(X^t;Y_t\mid Y^{t-1}).
 \label{eq:DI}
\end{equation}
The causal channel law is $W(y^n\Vert x^n,s_0)=\prod_tW_Y(y_t\mid s_{t-1},x_t)$. A deterministic feedback encoder satisfies $I(M;Y^n)=I_n$; a randomized encoder satisfies $I(M;Y^n)\le I_n$. These facts follow by inserting $X^t$ in the conditional chain rule and using the causal channel law. Thus a finite-block bound on $I_n$ is an operational converse without a stationarity assumption.

\section{Inverse design by dynamic measure matching}
\label{sec:matching}

Given a desired subchannel and causal codes achieving rate $C_\star$, the design problem is to specify the unused state--input rows so that the complete channel has capacity $C_\star$. A converse
for this purpose must cover every state that any physical input can
reach, including states outside the desired process.

The identity below separates the loss due to action choices from the divergence of
the output process. Corollary~\ref{cor:inverse} then gives the inverse
completion criterion.

\subsection{Operational design criterion}

\begin{theorem}[Exact dual--Bellman gap]
\label{thm:exactgap}
Let \eqref{eq:kernel}--\eqref{eq:domination} hold. Fix a real $\rho$, a potential $v$ on $\mathcal Z_{\mathrm{all}}$, and a nonnegative multiplier $\nu$ for the resource cost. Set
\begin{align}
 r_T(z,x)&=D(W_Y(\cdot\mid s,x)\|T(\cdot\mid q)),\label{eq:reward}\\
 \Delta_\nu(z,x)&=\rho+v(z)-r_T(z,x)+\nu b(s,x)\nonumber\\
                  &\quad-\mathbb E[v(Z^+)\mid z,x].\label{eq:residual}
\end{align}
For every causal input strategy and every blocklength,
\begin{align}
 &n\rho+v(z_0)-\mathbb E v(Z_n)
       +\nu\mathbb E\sum_{t=1}^n b(S_{t-1},X_t)-I_n\nonumber\\
 &\quad=\mathbb E\sum_{t=1}^n\Delta_\nu(Z_{t-1},X_t)
             +D(P_{Y^n}\|T_{Y^n}).
 \label{eq:exactgap}
\end{align}
If $\Delta_\nu\ge0$ throughout the all-policy closure, then
\begin{equation}
 C_{\mathrm{fb}}(B)\le\rho+\nu B.
 \label{eq:costupper}
\end{equation}
For $\nu=0$, this gives $C_{\mathrm{fb}}\le\rho$.
\end{theorem}

\begin{IEEEproof}
Unifilarity and the known initial state give
$P(Y_t\mid X^t,Y^{t-1})=W_Y(\cdot\mid S_{t-1},X_t)$. The conditional divergence identity yields
\begin{align}
 \mathbb E r_T(Z_{t-1},X_t)
  ={}&I(X^t;Y_t\mid Y^{t-1})\nonumber\\
    &+\mathbb E D(P_{Y_t\mid Y^{t-1}}\|T(\cdot\mid Q_{t-1})).
 \label{eq:dualidentity}
\end{align}
Summing and applying the relative-entropy chain rule gives
\begin{equation}
 \mathbb E\sum_{t=1}^n r_T(Z_{t-1},X_t)
   =I_n+D(P_{Y^n}\|T_{Y^n}).
 \label{eq:pathdual}
\end{equation}
On the other hand, summing \eqref{eq:residual} telescopes the potential terms:
\begin{align}
 \mathbb E\sum_{t=1}^n r_T
 ={}&n\rho+v(z_0)-\mathbb E v(Z_n)
      +\nu\mathbb E\sum_{t=1}^n b\nonumber\\
    &-\mathbb E\sum_{t=1}^n\Delta_\nu.
\end{align}
This proves \eqref{eq:exactgap}. Under a budget $B$, nonnegativity of the two terms on the right side gives $I_n\le n(\rho+\nu B)+\Span(v)$. Fano's inequality and $I(M;Y^n)\le I_n$ prove \eqref{eq:costupper}.
\end{IEEEproof}

The upper bound is the finite-state form of the duality method in
\cite{HuleihelSabagAharoniPermuter2025}. Retaining both slack terms
identifies the two matching conditions. Bellman equality removes the
local action residual; matching the conditional output process removes
the remaining divergence. An operational lower bound is provided by a
coding construction, as specified in the following criterion. This is a core result for inverse design.

\begin{corollary}[Inverse matching principle]
\label{cor:inverse}
Suppose an explicitly constructed sequence of feedback codes on a desired subchannel achieves rate $C_\star$ at budget $B$. If \eqref{eq:domination} and $\Delta_\nu\ge0$ hold for every physical input on the enlarged channel, with $C_\star=\rho+\nu B$, then $C_{\mathrm{fb}}(B)=C_\star$. If the desired codes have zero error, then $C_{0,\mathrm{fb}}(B)=C_\star$ as well.
\end{corollary}
\begin{IEEEproof}
The desired codes give the lower bound and Theorem~\ref{thm:exactgap} gives the upper bound.
\end{IEEEproof}

This criterion permits the unused channel rows to vary while the
desired code and its observation law remain fixed. Every row of the
completion must satisfy the converse at each compatible receiver
context. In particular, a row unused by the desired code can open new
physical trajectories; those trajectories must also lie in the checked
closure and satisfy support domination. Section~\ref{sec:synthesis}
turns these requirements into row-design constraints. The resulting
freedom in the unused rows also explains why optimal signaling can be
compatible with several channel mechanisms, as developed in
Section~\ref{sec:interpretation}.

\subsection{Behavior of every capacity-approaching code}

Once the desired rate equals capacity, a uniform positive residual on selected state--action pairs forces every reliable capacity-approaching sequence of codes to use those pairs with vanishing expected frequency. The physical actions remain available; their information penalty limits how often a good code can use them.

Let $\mathcal P\subseteq\mathcal Z_{\mathrm{all}}\times\mathcal X$ be the selected state--action pairs to be penalized and let
$N_{\mathrm p}(n)=\sum_{t=1}^n\mathbf1\{(Z_{t-1},X_t)\in\mathcal P\}$.

\begin{theorem}[Quantitative endogenous constraints]
\label{thm:stability}
Assume $\Delta_\nu\ge0$ everywhere and $\Delta_\nu\ge\delta>0$ on $\mathcal P$. For every code of expected average cost at most $B$,
\begin{align}
 (1-P_e^{(n)})\log_2 M_n
 \le{}&n(\rho+\nu B)+\Span(v)+1\nonumber\\
      &-\delta\,\mathbb E N_{\mathrm p}(n)
       -D(P_{Y^n}\|T_{Y^n}).
 \label{eq:codegap}
\end{align}
If reliable code rates tend to $C_\star=\rho+\nu B$, then
\begin{equation}
 \frac{\mathbb E N_{\mathrm p}(n)}n\to0,
 \quad \frac1nD(P_{Y^n}\|T_{Y^n})\to0.
 \label{eq:nearcapacity}
\end{equation}
The violation density also converges to zero in probability. If $\nu>0$, the normalized expected budget slack tends to zero.
\end{theorem}

\begin{IEEEproof}
For a code satisfying the budget, rewrite \eqref{eq:exactgap} as
\begin{align}
 &nC_\star+v(z_0)-\mathbb E v(Z_n)-I_n\nonumber\\
 &\quad=\mathbb E\sum_t\Delta_\nu
       +D(P_{Y^n}\|T_{Y^n})
       +\nu\left(nB-\mathbb E\sum_t b\right).
 \label{eq:budgetgap}
\end{align}
Each term on the right is nonnegative. Fano's inequality gives
$I_n\ge(1-P_e^{(n)})\log_2M_n-1$, proving \eqref{eq:codegap}. Dividing \eqref{eq:budgetgap} by $n$ proves the limits. Markov's inequality gives
$\Pr\{N_{\mathrm p}(n)/n>\varepsilon\}\le\mathbb E N_{\mathrm p}(n)/(n\varepsilon)$.
\end{IEEEproof}

The normalized divergence statement is a rate statement. It does not assert convergence of the entire block distribution in total variation. The chain rule also shows that the mean squared one-step predictive error vanishes: Pinsker's inequality gives
\begin{equation}
 \frac1n\sum_{t=1}^n\mathbb E\left\|P_{Y_t\mid Y^{t-1}}
             -T(\cdot\mid Q_{t-1})\right\|_{\rm TV}^2\to0.
 \label{eq:predictive-tv}
\end{equation}
In fact \eqref{eq:predictive-tv} follows from \eqref{eq:nearcapacity} for any fixed test process of the form \eqref{eq:test}; no stationarity of the code is needed.

\subsection{Dynamic surprise and continuation values}

For one channel state and one context, the potential cancels. Dynamic matching becomes
\begin{equation}
 D(W_Y(\cdot\mid x)\|T)-\nu b(x)\le\rho,
 \label{eq:static}
\end{equation}
with equality on used inputs. Let $P_Y^\star$ denote an output law induced by a capacity-achieving input distribution. When $T=P_Y^\star$, \eqref{eq:static} is the channel-side Karush--Kuhn--Tucker (KKT) condition underlying Gastpar's measure matching \cite{GastparRimoldiVetterli2003}; it involves Bayesian surprise \cite{WuGeVarshney2018Surprising}. With memory, the matched quantity is instead
\begin{equation}
 \mathsf B_T(z,x)=r_T(z,x)+\mathbb E v(Z^+)-v(z).
 \label{eq:corrected}
\end{equation}
This is the expected dynamic surprise conditional on $(z,x)$. Its immediate term is the mean log-likelihood ratio under the physical row $W_Y(\cdot\mid s,x)$, and its continuation term is the expected change in $v$. It satisfies $\mathsf B_T(z,x)-\nu b(s,x)=\rho-\Delta_\nu(z,x)$. 

The refractory construction of Section~\ref{sec:refractory} demonstrates
the distinction: ordinary and premature recovery requests have the
same immediate readout, while the latter enters a shadow recovery path
with a different continuation value. That path determines its
information penalty.

Define the resource-adjusted action value
\begin{equation}
 \mathsf Q_T^\nu(z,x)=r_T(z,x)-\nu b(s,x)+\mathbb E[v(Z^+)\mid z,x].
 \label{eq:surprise-action-value}
\end{equation}
Its advantage is $\mathsf Q_T^\nu(z,x)-\rho-v(z)=-\Delta_\nu(z,x)$. For $\nu=0$, \eqref{eq:corrected} equals $\mathsf Q_T^0(z,x)-v(z)$. The continuation potential $v$ is defined up to an additive constant; the corrected increment and the residual are invariant under that choice.

Appendix~\ref{app:surprise-interpretation} gives the Bayesian
posterior-update interpretation under a matched predictive law and the
hitting-time representation of the continuation potential. These
interpretations explain the value of future information; the design
criterion itself follows from the operational gap identity.

For an inverse resource problem with $\nu>0$, equality requires
\begin{equation}
 \nu b(s,x)=r_T(s,q,x)+\mathbb E v(Z^+)-v(s,q)-\rho.
 \label{eq:inversecost}
\end{equation}
A physical cost $b(s,x)$ must satisfy this relation at every compatible $q$ where the action is used. It cannot be assigned a different value for each receiver context unless that context is itself part of the physical resource model.

\subsection{Pathwise surprise and its compensator}

The continuation correction also has a trajectory-level form. Let $(\mathcal F_t)$ be the filtration generated by the message, the encoder's private random seed, and the output history, with the seed independent of channel noise. Then $X_t$ and $S_{t-1}$ are $\mathcal F_{t-1}$-measurable. Let $L_t$ be the realized log-likelihood ratio and $J_t$ the dynamic surprise after deducting the resource term:
\begin{align}
 L_t&=\log_2\frac{W_Y(Y_t\mid S_{t-1},X_t)}{T(Y_t\mid Q_{t-1})},\label{eq:realized-surprise}\\
 J_t&=L_t-\nu b(S_{t-1},X_t)+v(Z_t)-v(Z_{t-1}).
 \label{eq:realized-correction}
\end{align}
Only positive-probability realized outputs enter these expressions. Dynamic surprise is a signed, reference-dependent increment; the matching condition controls its conditional mean.

\begin{proposition}[Pathwise dynamic matching]
\label{prop:pathwise-surprise}
Under the support and finiteness assumptions of Theorem~\ref{thm:exactgap},
\begin{equation}
 \mathbb E[J_t\mid\mathcal F_{t-1}]
 =\rho-\Delta_\nu(Z_{t-1},X_t).
 \label{eq:conditional-surprise}
\end{equation}
Writing the causal channel likelihood as
$W(Y^n\Vert X^n,s_0)=\prod_{t=1}^nW_Y(Y_t\mid S_{t-1},X_t)$, we have
\begin{align}
 \sum_{t=1}^nJ_t
 ={}&\log_2\frac{W(Y^n\Vert X^n,s_0)}{T_{Y^n}(Y^n)}
       -\nu\sum_{t=1}^n b(S_{t-1},X_t)\nonumber\\
    &+v(Z_n)-v(z_0).
 \label{eq:pathwise-surprise}
\end{align}
Moreover,
\begin{equation}
 \mathcal M_n=\sum_{t=1}^n\{J_t-\rho+\Delta_\nu(Z_{t-1},X_t)\}
 \label{eq:surprise-martingale}
\end{equation}
is a martingale. Thus the sum of the conditional means in \eqref{eq:conditional-surprise} is the predictable compensator of $\sum_{t=1}^n J_t$. If $\Delta_\nu\ge0$, then $\sum_{t=1}^n(J_t-\rho)$ is a supermartingale; it is a martingale under a policy using only zero-residual actions.
\end{proposition}
\begin{IEEEproof}
Conditional on $\mathcal F_{t-1}$, the law of $Y_t$ is the selected channel row, so $\mathbb E[L_t\mid\mathcal F_{t-1}]=r_T(Z_{t-1},X_t)$. Substituting the residual definition gives \eqref{eq:conditional-surprise}. Summing \eqref{eq:realized-correction} telescopes the potential differences and proves \eqref{eq:pathwise-surprise}. The increments in \eqref{eq:surprise-martingale} have conditional mean zero. They are integrable because the channel and test laws are fixed on finite sets and the test law dominates every reachable row. The supermartingale assertion follows from the sign of the residual.
\end{IEEEproof}

The numerator in \eqref{eq:pathwise-surprise} is a causal likelihood, which generally differs from $P(Y^n\mid X^n)$ in the presence of feedback. Dynamic matching controls the conditional drift of pathwise surprise; a bound on each realized increment requires additional structure. Section~\ref{sec:refractory} gives a channel for which the martingale vanishes identically at a particular design parameter. On a matched predictive process, the log updates are Bayesian updates; off that process, the same identities compare channel trajectories with the fixed converse measure.

\section{Convex synthesis of channel rows}
\label{sec:synthesis}

Corollary~\ref{cor:inverse} reduces a channel completion to selecting
rows that preserve an upper bound already attained by the desired
subchannel. Take the desired rows and any other fixed rows as given.
We now optimize the remaining rows with the state-update topology,
test process, potential, resource law, allowed supports, candidate
bound $\rho$, and multiplier $\nu$ fixed. Selecting these ingredients
remains part of the inverse problem; the convexity result below
conditions on them.

One physical row $W_Y(\cdot\mid s,x)$ must satisfy the inequalities at
every receiver context compatible with $s$. The context indexes these
constraints, while the row is shared. The design must also specify a set of state--context pairs that contains the initial pair and is closed under every allowed transition.

Fix a finite design set $\mathcal Z_\circ$ containing the initial pair. For each row to be designed, choose an allowed output set $\mathcal A_{s,x}$ such that, for every $(s,q)\in\mathcal Z_\circ$ and $y\in\mathcal A_{s,x}$,
\begin{equation}
 T(y\mid q)>0,\quad
 (f(s,x,y),g(q,y))\in\mathcal Z_\circ.
 \label{eq:designsupport}
\end{equation}
All fixed rows satisfy the same closure and support conditions. Every trajectory under every feasible row choice therefore remains in $\mathcal Z_\circ$. The states and contexts to be checked can be chosen before the row probabilities are optimized; assigning zero probability to an allowed output preserves this closure.

For a probability row $w$ on $\mathcal A_{s,x}$, define
\begin{align}
 c_q(y)&=v(f(s,x,y),g(q,y))-v(s,q)-\nu b(s,x),\label{eq:cq}\\
 B_q(w)&=D(w\|T(\cdot\mid q))+\sum_yw(y)c_q(y).
 \label{eq:Bq}
\end{align}
For the candidate physical row $w$, $B_q(w)$ is expected dynamic surprise minus $\nu b(s,x)$; its Bellman residual at context $q$ is $\rho-B_q(w)$. Here $T$ remains normalized on the full output alphabet; the row $w$ is zero outside $\mathcal A_{s,x}$.

\begin{proposition}[Architecture-conditioned synthesis]
\label{prop:convex}
For fixed $(f,g,T,v,\rho,\nu,b)$ and allowed supports, the program
\begin{align}
 \text{maximize}\quad &\delta\nonumber\\
 \text{over}\quad &\{w_{s,x}\},\ \delta\nonumber\\
 \text{subject to}\quad &B_q(w_{s,x})+\delta\le\rho
 \quad\text{for all compatible }q
 \label{eq:convexdesign}
\end{align}
for the rows to be penalized is convex. Additional closed convex physical constraints on the rows preserve convexity. Feasibility with $\delta>0$, together with nonnegative residuals for every fixed row, tightness on the desired rows, and an achievable rate $\rho+\nu B$ at budget $B$, gives capacity equality by Corollary~\ref{cor:inverse} and the quantitative behavioral restriction of Theorem~\ref{thm:stability}.
\end{proposition}
\begin{IEEEproof}
Relative entropy is convex in its first argument, and the continuation term is affine. The constraints in \eqref{eq:convexdesign} are convex, the objective is linear, and \eqref{eq:designsupport} guarantees that all reachable pairs are checked.
\end{IEEEproof}

When only the output support restricts a row, we can maximize its smallest Bellman residual across the compatible receiver contexts. The solution is an exponential tilt of a geometric mixture of their test laws. Write
$\mathcal P(\mathcal A)$ for the simplex of probability distributions
on a finite set $\mathcal A$.

\begin{theorem}[Context-robust Gibbs row]
\label{thm:gibbs}
Fix a physical pair $(s,x)$ for which $\mathcal Q_s=\{q:(s,q)\in\mathcal Z_\circ\}\ne\varnothing$. Assume \eqref{eq:designsupport}, a nonempty common output set $\mathcal A$, and no row constraints beyond $w\in\mathcal P(\mathcal A)$. Define
\begin{align}
 K_\eta(y)&=2^{\sum_{q\in\mathcal Q_s}\eta_q\{\log_2T(y\mid q)-c_q(y)\}},\label{eq:Keta}\\
 Z(\eta)&=\sum_{y\in\mathcal A}K_\eta(y),
 \quad \eta\in\mathcal P(\mathcal Q_s).
 \label{eq:Zeta}
\end{align}
The largest common residual of this row is
\begin{equation}
 \delta^\star
 =\rho-\min_w\max_{q\in\mathcal Q_s}B_q(w)
 =\rho+\min_{\eta\in\mathcal P(\mathcal Q_s)}\log_2 Z(\eta).
 \label{eq:gibbsvalue}
\end{equation}
If $\eta^\star$ minimizes the last expression, a minimizing physical row is
\begin{equation}
 w^\star(y)=\frac{K_{\eta^\star}(y)}{Z(\eta^\star)}.
 \label{eq:gibbsrow}
\end{equation}
For one context $q$, this reduces to
\begin{equation}
 w^\star(y)=
 \frac{T(y\mid q)2^{-v(f(s,x,y),g(q,y))}}
 {\sum_{u\in\mathcal A}T(u\mid q)2^{-v(f(s,x,u),g(q,u))}}.
 \label{eq:singlecontext}
\end{equation}
\end{theorem}

\begin{IEEEproof}
For any $w$, $\max_qB_q(w)=\max_\eta\sum_q\eta_qB_q(w)$. This function is continuous and convex in $w$ on a compact simplex, and affine in $\eta$. The minimax theorem \cite{Sion1958} gives
\begin{equation}
 \min_w\max_\eta\sum_q\eta_qB_q(w)
   =\max_\eta\min_w\sum_q\eta_qB_q(w).
\end{equation}
For fixed $\eta$, direct expansion yields
\begin{equation}
 \sum_q\eta_qB_q(w)
   =D\left(w\middle\|\frac{K_\eta}{Z(\eta)}\right)-\log_2Z(\eta).
 \label{eq:gibbsvariational}
\end{equation}
Its unique minimizer is $K_\eta/Z(\eta)$, so the minimax value is $\max_\eta[-\log_2Z(\eta)]$. This proves \eqref{eq:gibbsvalue}. At an optimal $\eta^\star$, the unique inner minimizer is the primal minimax optimizer: a saddle point exists, and strict convexity makes its minimizing row unique. Equation~\eqref{eq:singlecontext} follows because the terms $v(s,q)+\nu b(s,x)$ cancel in normalization.
\end{IEEEproof}

The mixing weights $\eta$ enforce compatibility across receiver
contexts. With one context the row is an exponentially tilted test
distribution; with several contexts it is a tilt of their geometric
mixture. Independently chosen rows for the different contexts would
have to agree to define the same physical channel. The optimization in
\eqref{eq:gibbsvalue} is convex in $\eta$ because $\log Z$ is a
log-sum-exp of affine functions. A positive $\delta^\star$ gives the
required strict margin for this row. A nonpositive value means that
these fixed design ingredients do not provide such a margin.

\subsection{Robustness on a fixed support}

A strict Bellman gap permits small changes in unused channel rows without changing capacity, provided their output supports and the desired subchannel remain fixed.

\begin{corollary}[Support-preserving stability of capacity]
\label{cor:robustness}
Suppose a channel satisfies \eqref{eq:domination} and $\Delta_\nu\ge0$ on its all-policy closure, has an explicit desired code attaining $C_\star=\rho+\nu B$, and all changed rows are unused by that code and have residual at least $\delta>0$ at every compatible context. Keep $f,g,T,v,b$, the other rows, and the exact support of each changed row fixed. Then all sufficiently small changes in those rows preserve residual at least $\delta/2$ and preserve $C_{\mathrm{fb}}(B)=C_\star$. Zero-error capacity is also preserved when the desired code has zero error.
\end{corollary}
\begin{IEEEproof}
Fixed supports and updates leave the all-policy closure unchanged. On each fixed support face, $D(w\|T(\cdot\mid q))$ and the expected successor potential are continuous in $w$, including at zero coordinates, because the test law is positive on every allowed output. Finitely many strict inequalities therefore persist in a common total-variation neighborhood. The desired code is unchanged, and the converse remains valid.
\end{IEEEproof}

Note that the restriction on support is important. Adding a previously impossible output can enter an unchecked product state or encounter a zero of $T$. Parameterized constructions can allow larger changes, including changes in tight autonomous rows, by adjusting the continuation potential. Section~\ref{sec:refractory} gives such a family explicitly.

\section{Constructive inverse design from constrained processes}
\label{sec:constraints}

Let us consider the setting where the desired family of signaling paths are finite-state constrained systems, cf. the trapdoor channel's signaling scheme \cite{PermuterCuffVanRoyWeissman2008}. The constrained graph's Perron root and positive eigenvector provide a test process and a continuation potential, the ingredients held fixed in Section~\ref{sec:synthesis}. We use them to construct a channel in
which every physical input remains available and the prescribed
constrained entropy is the capacity.

Let $G=(V,E)$ be a finite irreducible, or strongly connected, right-resolving graph with a fixed initial vertex $i_0$. Right-resolving means that different outgoing edges have different labels. Edge identities themselves can be used as labels. For an edge $e:i\to j$, write $\tail(e)=i$ and $\head(e)=j$. Let
\begin{equation}
 A_{ij}=|\{e:i\to j\}|,\quad Ar=\lambda r,\quad r>0.
 \label{eq:PF}
\end{equation}
Here $A$ is the adjacency matrix, $\lambda$ is its Perron root, and $r$ is a positive right eigenvector. Right-resolvingness gives a bijection between paths from $i_0$ and their label strings. Let $e_i$ be the coordinate vector at vertex $i$ and $\mathbf1$ the all-ones vector. The number of length-$n$ paths from $i$ is $N_i(n)=e_i^{\mathsf T}A^n\mathbf1$, and
\begin{equation}
 \frac{\lambda^n r_i}{\max_jr_j}\le N_i(n)
 \le\frac{\lambda^n r_i}{\min_jr_j}.
 \label{eq:countbounds}
\end{equation}
These bounds hold at every blocklength and give constrained entropy $\rho=\log_2\lambda$, without an aperiodicity assumption \cite{LindMarcus2021,MarcusRothSiegel1998}.

\subsection{The Perron potential}

The Shannon--Parry measure has transition probabilities \cite{Parry1964}
\begin{equation}
 \pi^\star(e\mid i)=\frac{r_j}{\lambda r_i},\quad e:i\to j,
 \quad h(i)=\log_2r_i.
 \label{eq:parry}
\end{equation}
Let $\Pi_{i_0}^{(n)}$ be its law on length-$n$ paths from $i_0$. For an edge path $E^n$, let $V_t$ be the vertex reached after edge $E_t$, with $V_0=i_0$.

\begin{proposition}[Perron--Bellman identity and path entropy]
\label{prop:perron}
For each $i$,
\begin{equation}
 \rho+h(i)=\max_{\pi(\cdot\mid i)}
 \left\{H(\pi)+\sum_{e:i\to j}\pi(e\mid i)h(j)\right\}.
 \label{eq:PFbellman}
\end{equation}
The unique edge-law maximizer is \eqref{eq:parry}. For every random legal path $E^n$ from $i_0$, including one selected using arbitrary causal side observations,
\begin{align}
 H(E^n)+D(P_{E^n}\|\Pi_{i_0}^{(n)})
   =n\rho+h(i_0)-\mathbb E h(V_n).
 \label{eq:PFexact}
\end{align}
In particular, $H(E^n)\le n\rho+\Span(h)$.
\end{proposition}
\begin{IEEEproof}
The log-sum variational identity gives
$\max_\pi\{H(\pi)+\sum_e\pi(e)h(\head(e))\}
\allowbreak=\log_2\sum_{e:i\to j}r_j\allowbreak=\log_2(Ar)_i$.
For a legal path, \eqref{eq:parry} telescopes:
\begin{equation}
 -\log_2\Pi_{i_0}^{(n)}(e^n)
   =n\rho+h(i_0)-h(i_n).
 \label{eq:parrypath}
\end{equation}
Taking expectation under any $P_{E^n}$ proves \eqref{eq:PFexact}.
\end{IEEEproof}

The maximizing row in \eqref{eq:PFbellman} need not be uniform: its probabilities account for the different numbers of continuations available from the successor vertices. These are classical maximum-entropy identities. Their use here is operational: arbitrary feedback dependence does not invalidate a bound that holds for every probability law on the legal path set. The potential $h(i)$ records the state-dependent abundance of future paths, while the local edge surprisal satisfies
\begin{equation}
 -\log_2\pi^\star(e\mid i)+h(j)-h(i)=\rho.
 \label{eq:edge-matching}
\end{equation}

\subsection{A universal soft realization}

Retain a controlled mode in which each desired edge input is emitted
exactly. Complete the channel by letting an undesired input initiate an
autonomous excursion drawn from the target Shannon--Parry measure. The encoder then
relinquishes the choice of the emitted path until control returns. The
following construction makes the entropy of that excursion its exact
Bellman penalty.

Let $P^\star$ be the vertex transition matrix of \eqref{eq:parry}, and assume $\rho>0$. Choose an integer $L\ge1$. The channel $\mathsf{Soft}(G,L)$ has input alphabet $E\cup\{\bot\}$, where $\bot$ is an additional input symbol, output alphabet $E$, and state $(i,m)$. Here $i$ is the current graph vertex and $m\in\{0,\ldots,L-1\}$ counts the autonomous emissions remaining before control returns. The initial channel state is $(i_0,0)$. At $m=0$, the encoder can select the next edge. A legal edge input $e:i\to j$ produces $Y=e$ and state $(j,0)$. Any other input produces $Y=e\sim\pi^\star(\cdot\mid i)$ and state $(\head(e),L-1)$. In a mode $m>0$, every input produces $e\sim\pi^\star(\cdot\mid i)$ and state $(\head(e),m-1)$.

An excursion comprises $L$ autonomous emissions including its initiating use. All physical inputs remain available. The output identifies the graph vertex, so take $Q=i$ and $T(e\mid i)=\pi^\star(e\mid i)$. The physical mode remains separate from this output context.

\begin{theorem}[Universal soft realization]
\label{thm:universal}
For every such graph, the channel $\mathsf{Soft}(G,L)$ has
\begin{equation}
 C_{0,\mathrm{fb}}=C_{\mathrm{fb}}=\rho=\log_2\lambda.
 \label{eq:universal-cap}
\end{equation}
With
\begin{equation}
 v(i,m)=(P^\star)^m h(i)-m\rho,
 \label{eq:universal-v}
\end{equation}
legal and autonomous rows are Bellman-tight. Every violating controlled-mode input has residual
\begin{align}
 \delta_L(i)
 &=L\rho+h(i)-(P^\star)^Lh(i)\nonumber\\
 &=H_{\pi^\star}(E^L\mid V_0=i).
 \label{eq:universal-entropy-gap}
\end{align}
In particular, if $L\rho>\Span(h)$, the realization is soft, the residual is uniformly positive, and every reliable capacity-approaching code has vanishing expected frequency of violating controlled-mode inputs.
\end{theorem}

\begin{IEEEproof}
For a legal controlled edge $e:i\to j$, the reward is $-\log_2\pi^\star(e\mid i)$, so \eqref{eq:edge-matching} gives
$r_T+v(j,0)=\rho+v(i,0)$.
For $m>0$, the physical row equals $T$, its reward is zero, and
\begin{equation}
 \sum_jP^\star_{ij}v(j,m-1)=\rho+v(i,m).
\end{equation}
A violating controlled-mode row also has zero reward, and its expected successor potential is
$(P^\star)^Lh(i)-(L-1)\rho$. Subtraction gives the first line of \eqref{eq:universal-entropy-gap}. The second follows by applying \eqref{eq:PFexact} to the Parry process itself. Thus the residual is nonnegative for every $L$, and is bounded below by $L\rho-\Span(h)$. The all-policy closure has $Q=i$; every output is supported by the Parry test row. Theorem~\ref{thm:exactgap} gives the upper bound.

For the lower bound, a length-$n$ legal edge path is sent deterministically in controlled mode and appears unchanged at the output. There are $N_{i_0}(n)$ such messages. The code is zero error and does not require feedback. This proves \eqref{eq:universal-cap}; Theorem~\ref{thm:stability} proves the frequency claim. If every $\delta_L(i)>0$, the autonomous $L$-edge law is not a point mass from any vertex. For any desired initial edge, select a fixed legal continuation. That desired trajectory is deterministic, whereas the violating action gives the nondegenerate autonomous path under the same continuation. The actions are behaviorally distinct, as required for softness.
\end{IEEEproof}

The capacity value also has a direct counting proof: every output is a legal graph path, so $I(M;Y^n)\le\log_2N_{i_0}(n)$. The Bellman argument adds the entropy penalty \eqref{eq:universal-entropy-gap}. Since at most $L-1$ autonomous uses follow each initiating violation, their density also vanishes near capacity when the gap is positive.

For a right-resolving graph, edge identities may be replaced by labels. Given the initial vertex and successive labels, the receiver reconstructs the same path. A missing label initiates the autonomous excursion. In particular, binary right-resolving constraints admit binary-input, binary-output realizations at vertices with a missing binary label. A spare symbol $\bot$ can be retained when a violation is to be provided at every vertex.

\subsection{A forbidden-word example}

Most common constrained systems have run-length constraints, but one can consider other possibilities. To forbid $101$, track the longest suffix of the transmitted sequence that is a proper prefix of $101$. The resulting states are $\epsilon,1,10$, where $\epsilon$ denotes the empty suffix, and the adjacency matrix is
\begin{equation}
 A=\begin{pmatrix}1&1&0\\0&1&1\\1&0&0\end{pmatrix}.
 \label{eq:A101}
\end{equation}
Its Perron root solves $\lambda^3-2\lambda^2+\lambda-1=0$ and equals $1.754877666247$. A right eigenvector is
\begin{equation}
 r=\bigl(\lambda,(\lambda-1)^{-1},1\bigr)^{\mathsf T}.
\end{equation}
The polynomial gives $\lambda(\lambda-1)^2=1$, which verifies all three eigenvector equations and $1<(\lambda-1)^{-1}<\lambda$. Thus $\Span(h)=\log_2\lambda$. Taking $L=2$ gives a binary soft realization with capacity $0.811370462752$ bits/use. The missing $1$-input at suffix $10$ initiates the excursion; its residual is $0.9858799038$ bits. The other vertex gaps of the edge-alphabet realization are $1.9717598077$ and $1.5476739951$ bits. The construction applies directly to the constrained graph without requiring a renewal representation.

\subsection{Enumeration and other realizations}

State-conditioned path counts provide another way to obtain the design
ingredients. Appendix~\ref{app:enumeration} relates their generating
functions to the Perron--Frobenius eigenvector. The scalar growth rate determines
$\rho$, while the state-resolved residues determine $h$ up to an
additive constant. The growth rate alone is therefore insufficient to
specify the continuation values needed for synthesis.

For a positive integer $d$, the binary $(d,\infty)$ constraint requires
at least $d$ zeros between successive ones. Let $N_d(n)$ count the legal
length-$n$ strings and let $F_d(z)=\sum_{n\ge0}N_d(n)z^n$. The
Goulden--Jackson derivation in Appendix~\ref{app:enumeration} gives
\eqref{eq:RLLGF} and the recurrence
\begin{equation*}
 N_d(n)=N_d(n-1)+N_d(n-d-1),\quad n\ge d+1,
\end{equation*}
with $N_d(n)=n+1$ for $0\le n\le d$. These counts will give the
coding rate of the refractory construction in
Section~\ref{sec:refractory}.

A further realization criterion applies when every physical request
admits a causal projection onto a legal graph edge.
Proposition~\ref{prop:quotient}, stated and proved in
Appendix~\ref{app:effective-actions}, combines that projection with
state-compatible input selectors and finite-delay decoding. Under its
conditional-independence assumption, it gives an exact path-counting
converse for all physical strategies; the selectors and observer provide
a separate zero-error construction.

\subsection{Resource-weighted target processes}

The same construction can begin with a stationary edge-cost constraint. For $c(e)\ge0$ and $\theta\ge0$, let
\begin{equation}
 A_\theta(i,j)=\sum_{e:i\to j}2^{-\theta c(e)},\quad
 A_\theta r_\theta=\Lambda_\theta r_\theta.
\end{equation}
Set $P(\theta)=\log_2\Lambda_\theta$, $h_\theta=\log_2r_\theta$, and
\begin{equation}
 \pi_\theta(e\mid i)=
 \frac{2^{-\theta c(e)}r_\theta(j)}{\Lambda_\theta r_\theta(i)}.
 \label{eq:tilted}
\end{equation}
Then
\begin{equation}
 -\log_2\pi_\theta(e\mid i)
 =P(\theta)+\theta c(e)+h_\theta(i)-h_\theta(j).
 \label{eq:weighted-edge}
\end{equation}
This is dynamic cost matching on graph paths. Appendix~\ref{app:weighted} gives the resource-constrained realization and its fixed-budget coding proof. As in probabilistic measure matching~\cite{GastparRimoldiVetterli2003}, the construction establishes optimality at a prescribed operating point. Here the autonomous transition law depends on $\theta$, so varying $\theta$ generally produces a family of channels rather than different operating points of one fixed channel.

\section{A parameterized family of soft refractory channels}
\label{sec:refractory}

We now design a compact channel with a noisy activation response. The prescribed behavior activates a ready system and then allows $d$ recovery uses before another activation. The recovery outputs resolve the initial readout ambiguity. Premature requests remain physically available: they give the same immediate output as an ordinary recovery request while changing the subsequent conformation.

The remaining design freedom is the autonomous response following such a request. A shadow recovery chain leads to a state $J$, whose output is one with probability $a$. Matching its continuation value produces the common action residual $H_2(a)-a$. We show that its nonnegative range is exactly the capacity plateau and that Shannon and zero-error feedback capacities differ beyond the plateau for $a<1$. The Gibbs row $a=1/3$ maximizes the residual and makes dynamic surprise equal its conditional mean on every possible transition, giving an exact trajectory-wise likelihood identity and a finite-block converse. Positive values of the residual also bound the information available about the unknown mechanism. We first construct the desired code and its output test process, then select the off-policy dynamics through the Bellman conditions.

\subsection{Prescribed signaling and channel architecture}

Fix $d\ge2$ and $a\in[0,1]$. We denote the reduced soft refractory channel by $\mathsf{SRT}_d(a)$. It has binary requests $U$, binary differential observations $D$, and conformation
\begin{equation}
 C\in\{R,F_d,\ldots,F_1,S_{d-1},\ldots,S_1,J\}.
 \label{eq:conformations}
\end{equation}
The observation $D=1$ records a change in the physical readout from the previous use. The desired behavior uses the ready state $R$ and normal recovery states $F_d,\ldots,F_1$. The additional states $S_{d-1},\ldots,S_1,J$ determine the response to a premature request. There are $2d+1$ conformations in total. Set $F_0=R$ and $S_0=J$. The initial conformation is $R$. The complete kernel is
\begin{equation}
\begin{array}{c|c|c|c}
 C&U&D&C^+\\\hline
 R&0&0&R\\
 R&1&\Bern(1/2)&F_d\\
 F_d&0&1&F_{d-1}\\
 F_d&1&1&S_{d-1}\\
 F_k,
 \ 1\le k<d&0&0&F_{k-1}\\
 F_k,
 \ 1\le k<d&1&0&S_{k-1}\\
 S_k,
 \ 1\le k<d&0\text{ or }1&0&S_{k-1}\\
 J&0\text{ or }1&0\ (1-a)&R\\
 J&0\text{ or }1&1\ (a)&F_d.
\end{array}
\label{tab:channel}
\end{equation}
The last two rows specify the probabilities in parentheses. Every transition is unifilar. A request $U=1$ in $F_k$ is premature. It has the same immediate output as $U=0$ but enters the shadow chain, which ends at $J$. For $0<a\le1$, the two recovery requests have different future output laws under the all-zero continuation: ordinary recovery returns to $R$, whereas shadow recovery reaches $J$, where the next observation is $\operatorname{Bernoulli}(a)$. At $a=0$, follow either request in $F_k$ by the common continuation $0^{k-1}1$. Before its final request the two trajectories are at $R$ and $J$, respectively; that request has a fair differential output at $R$ and gives zero deterministically at $J$. Thus the realization is soft throughout $a\in[0,1]$.

A two-species label representation uses a retained bit $B$ and binary physical input and readout $X,Y$. Its full kernel is given in Appendix~\ref{app:physical-coordinates}. Write $\oplus$ for addition modulo two. The following coordinate relation connects that representation to the reduced requests and observations.

\begin{lemma}[Causal equivalence of the two representations]
\label{lem:coordinates}
Fix $(B_0,C_0)=(b_0,R)$ and let $Y_0=b_0$. The physical channel and \eqref{tab:channel} have the same feedback and zero-error feedback capacities. Their causal input strategies correspond through
\begin{equation}
 U_t=X_t\oplus B_{t-1},\quad D_t=Y_t\oplus Y_{t-1}.
 \label{eq:coordinates}
\end{equation}
The induced directed information is also preserved.
\end{lemma}
\begin{IEEEproof}
See Appendix~\ref{app:physical-coordinates}.
\end{IEEEproof}

\subsection{Desired paths and zero-error decoding}

The desired strategy uses $U=0$ in normal recovery and never enters shadow states. Let $A_t$ be its activation request. Then
\begin{equation}
 A_t=1\ \Longrightarrow\ A_{t+1}=\cdots=A_{t+d}=0.
 \label{eq:RLL}
\end{equation}
Every sequence with this property is implemented by a fixed reduced input sequence; Lemma~\ref{lem:coordinates} implements it physically with feedback.

\begin{lemma}[Finite-delay observer]
\label{lem:observer}
For every desired trajectory and $d\ge2$,
\begin{equation}
 A_t=1\quad\Longleftrightarrow\quad D_{t+1}^{t+d}=10^{d-1}.
 \label{eq:observer}
\end{equation}
Appending $d$ zero requests therefore decodes the information path with zero error.
\end{lemma}
\begin{IEEEproof}
An activation enters $F_d$, whose next $d$ outputs are $10^{d-1}$. Conversely, suppose $A_t=0$. If $A_{t+1}=0$, then the next conformation cannot be $F_d$, which is entered only by an activation on a desired trajectory; hence $D_{t+1}=0$. If $A_{t+1}=1$, then $D_{t+2}=1$. Either possibility excludes the indicated signature. A later activation cannot change its first symbol.
\end{IEEEproof}

Let $\lambda_d>1$ solve
\begin{equation}
 \lambda_d^{d+1}=\lambda_d^d+1,
 \quad \rho_d=\log_2\lambda_d.
 \label{eq:meru}
\end{equation}
The counts in \eqref{eq:RLLGF} give zero-error rate $\rho_d$ for every value of $a$. The maximum-entropy activation probability at $R$ is
\begin{equation}
 p_d=\lambda_d^{-(d+1)},\quad 1-p_d=\lambda_d^{-1}.
 \label{eq:activation}
\end{equation}
For example, the constraint Bellman equation with $h(F_k)=-k\rho_d$ is
\begin{equation}
 \rho_d=\max_{0\le u\le1}\{H_2(u)-d\rho_d u\}
       =\log_2(1+2^{-d\rho_d}),
\end{equation}
which gives both \eqref{eq:meru} and \eqref{eq:activation}. Its entropy rate is $H_2(p_d)/(1+dp_d)=\rho_d$. The denominator is the mean number of channel uses per decision in $R$: one for the request and $d$ additional recovery uses with probability $p_d$.

\subsection{Test process inherited from the desired behavior}

The desired activation law determines the reference output process before the off-policy parameter $a$ is selected. For the remainder of this section write $\lambda=\lambda_d$, $\rho=\rho_d$, $p=p_d$, and
\begin{equation}
 a_0=p/2,\quad \ell=\log_2(\lambda+1).
 \label{eq:testparams}
\end{equation}
The test parameter $a_0$ is distinct from the physical parameter $a$ at $J$.

Use the $d+3$ contexts $q_0,q_1,q_2,q_3,u_{d-1},\ldots,u_1$, with $u_0=q_0$. Their relevant edges are
\begin{align}
 g(q_0,0)&=q_1,&g(q_0,1)&=q_2,\nonumber\\
 g(q_1,0)&=q_1,&g(q_1,1)&=q_3,\nonumber\\
 g(q_2,1)&=u_{d-1},\nonumber\\
 g(q_3,0)&=u_{d-2},&g(q_3,1)&=u_{d-1},\nonumber\\
 g(u_k,0)&=u_{k-1}.&&
 \label{eq:Qgraph}
\end{align}
The unspecified edges can be assigned arbitrary successors; Lemma~\ref{lem:allpolicy} will show that no physical strategy traverses them. The test law is
\begin{align}
 T(1\mid q_0)&=a_0,& T(1\mid q_1)&=p,\nonumber\\
 T(1\mid q_2)&=1,& T(1\mid q_3)&=(\lambda+1)^{-1},\nonumber\\
 T(1\mid u_k)&=0.&&
 \label{eq:refractorytest}
\end{align}
Under the desired Markov action law, these are the actual conditional output laws. Starting from known $R$, observation zero gives the posterior
\begin{equation}
 q_1:\quad P(R)=\frac{1-p}{1-p/2},\quad
 P(F_d)=\frac{p/2}{1-p/2}.
 \label{eq:q1belief}
\end{equation}
Another zero returns the same posterior. At $q_1$, output one occurs with probability $p$ and produces
\begin{equation}
 q_3:\quad P(F_d)=\frac1{\lambda+1},\quad
 P(F_{d-1})=\frac\lambda{\lambda+1}.
 \label{eq:q3belief}
\end{equation}
Outputs from $q_3$ resolve the branch, after which the known recovery state counts down. These Bayes calculations explain the finite test graph. Under other policies its nodes are reference contexts, not assertions about the actual posterior.

\begin{figure*}[t]
\centering
\resizebox{0.98\textwidth}{!}{%
\begin{tikzpicture}[font=\scriptsize]
\begin{scope}
\node[ibnormal] (R) at (0,0) {$R$};
\node[ibnormal] (F2) at (2.8,.65) {$F_2$};
\node[ibnormal] (F1) at (5.6,0) {$F_1$};
\node[ibshadow] (S1) at (2.8,-1.85) {$S_1$};
\node[ibmatched] (J) at (5.6,-1.85) {$J$};
\draw[iblegal] (R) edge[loop left,looseness=5] node[left] {$0/0$} (R);
\draw[iblegal] (R) to[bend left=4] node[above,sloped,pos=.55] {$1/\Bern(1/2)$} (F2);
\draw[iblegal] (F2)--node[above,pos=.5] {$0/1$} (F1);
\draw[iblegal] (F1.north) .. controls (4.5,2.0) and (1.1,2.0) .. node[above,pos=.5] {$0/0$} (R.north);
\draw[ibpremature] (F2)--node[left,pos=.52] {$1/1$} (S1);
\draw[ibpremature] (F1)--node[right,pos=.52] {$1/0$} (J);
\draw[ibauto] (S1)--node[below,pos=.5] {$*/0$} (J);
\draw[ibauto] (J) to[bend right=15] node[right,pos=.5] {$a:1$} (F2);
\draw[ibauto] (J.south) .. controls (5.4,-3.55) and (-1.0,-3.55) .. node[below,pos=.45] {$(1-a):0$} (R.south);
\node[font=\scriptsize\bfseries,IBInk] at (2.55,-3.85) {(a) Physical conformation dynamics};
\end{scope}
\begin{scope}[xshift=8cm]
\node[ibnormal] (q0) at (0,0) {$q_0$};
\node[ibnormal] (q1) at (2.6,0) {$q_1$};
\node[ibnormal] (q2) at (0,-1.9) {$q_2$};
\node[ibnormal] (q3) at (2.6,-1.9) {$q_3$};
\node[ibnormal] (u1) at (5.2,-.95) {$u_1$};
\draw[ibarrow] (q0)--node[above] {$0$} (q1);
\draw[ibarrow] (q0)--node[left] {$1$} (q2);
\draw[ibarrow] (q1) edge[loop above,looseness=5] node[above] {$0$} (q1);
\draw[ibarrow] (q1)--node[right] {$1$} (q3);
\draw[ibarrow] (q3)--node[above,sloped] {$1$} (u1);
\draw[ibarrow] (q3)--node[above,sloped,pos=.57] {$0$} (q0);
\draw[ibarrow] (q2.south) .. controls (0,-3.55) and (5.2,-3.55) .. node[below,pos=.47] {$1$} (u1.south);
\draw[ibarrow] (u1.north) .. controls (5.2,2.0) and (0,2.0) .. node[above,pos=.55] {$0$} (q0.north);
\node[font=\scriptsize\bfseries,IBInk] at (2.6,-3.85) {(b) Output-recursive context};
\end{scope}
\end{tikzpicture}}
\caption{Soft refractory realization for $d=2$. (a) Desired actions are solid, premature actions dashed, and autonomous transitions dotted. Shadow states have dashed outlines, and the desensitized state has a double outline. Labels are request/output pairs $U/D$, except at $J$, where they are probability/output pairs; $*$ denotes either request. (b) Context edges are labeled by the differential output. The channel uses the free physical parameter $a$, whereas the test law uses the fixed value $a_0=p/2$. Completion edges with zero probability on the all-policy closure are omitted.}
\label{fig:refractory}
\end{figure*}
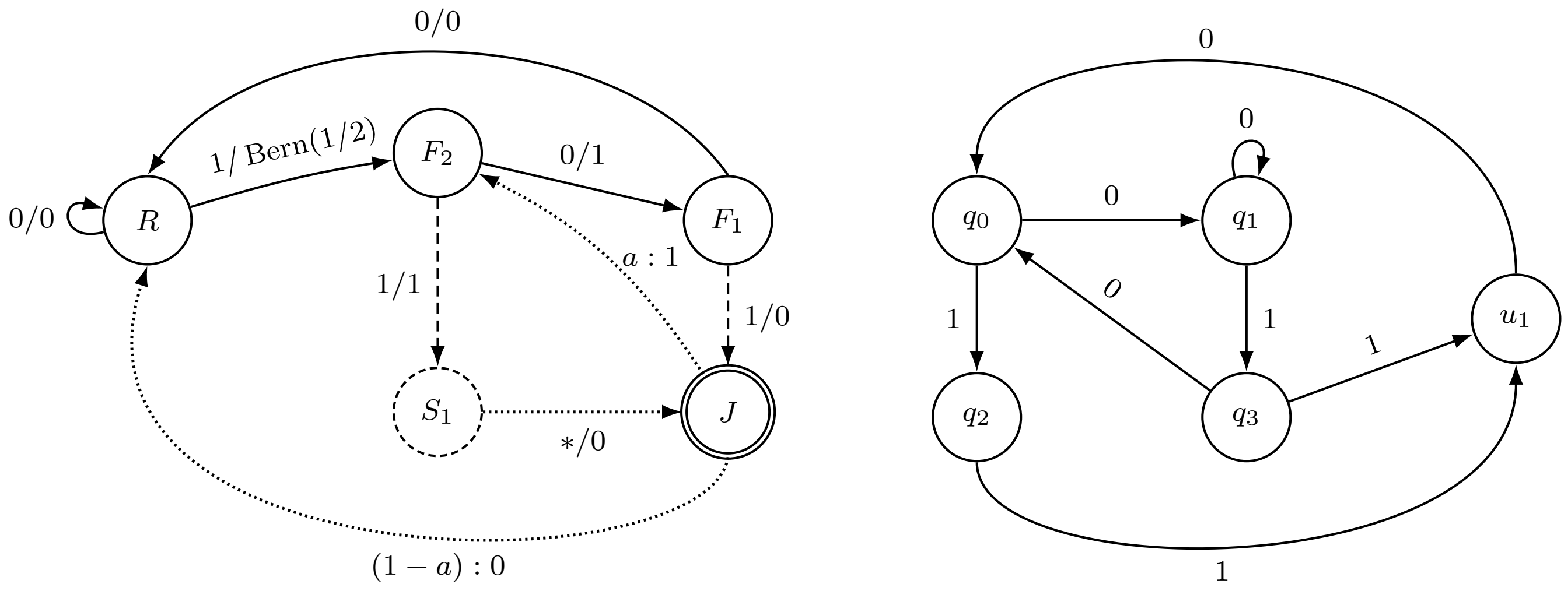

\subsection{Closure under all physical requests}

Let
\begin{align}
 \mathcal Z_{\rm N}={}&\{(R,q_0),(R,q_1),
 (F_d,q_1),(F_d,q_2),(F_d,q_3),\nonumber\\
 &\hspace{8mm}(F_{d-1},q_3)\}
 \cup\{(F_k,u_k):1\le k<d\},\label{eq:normalset}\\
 \mathcal Z_{\rm S}={}&\{(J,q_0),(S_{d-1},q_3)\}
 \cup\{(S_k,u_k):1\le k<d\}.
 \label{eq:shadowset}
\end{align}
These sets contain $2d+6$ pairs in total.

\begin{lemma}[All-policy closure and domination]
\label{lem:allpolicy}
For every $d\ge2$ and $a\in[0,1]$, the all-policy reachable set from $(R,q_0)$ is $\mathcal Z_{\rm N}\cup\mathcal Z_{\rm S}$. Every physical request at every pair in this set is dominated by \eqref{eq:refractorytest}.
\end{lemma}
The complete transition table and proof are given in Appendix~\ref{app:allpolicy}. The closure includes the shadow states reached by premature requests, so the converse based on this test process applies to every physical strategy. Appendix~\ref{app:physical-coordinates} gives the corresponding output-recursive test process for the physical readout $Y$.

\subsection{Inverse design of the shadow dynamics}

The test process defines the information reward against which the physical rows are designed. We choose the continuation values of the shadow states so that premature requests incur the same residual at every compatible receiver context. This makes the effect of their different future dynamics explicit.

Keep the desired subchannel and its test law fixed. Choose the following normal-state potential, which does not depend on $a$:
\begin{align}
 v(R,q_0)&=0,&v(R,q_1)&=\ell-1,\nonumber\\
 v(F_d,q_1)&=\ell,&v(F_d,q_2)&=-d\rho,\nonumber\\
 v(F_d,q_3)&=\ell-d\rho,&v(F_{d-1},q_3)&=\ell-d\rho,\nonumber\\
 v(F_k,u_k)&=-k\rho,&&1\le k<d.
 \label{eq:normalpotential}
\end{align}

At $J$, the actual row is $\Bern(a)$ and the test row is $\Bern(a_0)$. Its Bellman equality requires
\begin{align}
 v(J,q_0)
 ={}&D_{\rm B}(a\|a_0)-\rho
 +(1-a)(\ell-1)-ad\rho.\label{eq:Jrequired}
\end{align}
Expanding the binary divergence gives
\begin{align}
 D_{\rm B}(a\|a_0)
  ={}&-H_2(a)+a\{(d+1)\rho+1\}\nonumber\\
     &+(1-a)(\rho-\ell+1).
\end{align}
Define
\begin{equation}
 \kappa(a)=a-H_2(a),\quad \delta(a)=-\kappa(a),
 \label{eq:kappaa}
\end{equation}
and extend the potential by
\begin{align}
 v(J,q_0)&=\kappa(a),\nonumber\\
 v(S_{d-1},q_3)&=\ell-d\rho+\kappa(a),\nonumber\\
 v(S_k,u_k)&=-k\rho+\kappa(a).
 \label{eq:shadowpotential}
\end{align}

The row at $J$ is therefore Bellman-tight. The shadow-chain equalities follow by deterministic countdown. A premature action has the same immediate output and the same next context as the desired recovery action, but its successor potential is shifted by $\kappa(a)$. Its residual is therefore $-\kappa(a)$ at every compatible context.

Varying $a$ changes a tight autonomous row and shifts all shadow potentials by $\kappa(a)$. With the desired subchannel, $T$, and the normal potential fixed, the single-context row problem at $J$ minimizes
\begin{equation}
 G(a)=D_{\rm B}(a\|a_0)+(1-a)(\ell-1)-ad\rho.
 \label{eq:Jobjective}
\end{equation}
Theorem~\ref{thm:gibbs} gives weights proportional to
\begin{equation}
 (1-a_0)2^{-(\ell-1)}=\lambda^{-1},
 \quad a_0 2^{d\rho}=(2\lambda)^{-1}.
\end{equation}
Thus the designed row is $a=1/3$. Equivalently,
\begin{equation}
 \max_{a\in[0,1]}\delta(a)=\delta(1/3)=\log_2(3/2).
 \label{eq:maxgap}
\end{equation}
The derivative is $\delta'(a)=\log_2((1-a)/a)-1$, so the maximizer is unique. This optimization is over the displayed family with its normal-state potential fixed.

At $a=a_0$, the physical and test output laws at $J$ agree, so their divergence is zero. The capacity result below permits an interval of choices for this row. The maximum residual occurs at $a=1/3$, which minimizes the sum of the immediate divergence and the expected successor potential.

\subsubsection{Meru instances}

The growth factors $\lambda_2$ and $\lambda_3$ are, respectively, the second and fourth Meru constants associated with Pi\.{n}gala's \emph{Meru-prast\={a}ra} \cite{Varshney2008Meru}. These constants are the asymptotic growth factors of the corresponding diagonal-sum sequences. For $d=2$, $\lambda_2^3=\lambda_2^2+1$ gives the second Meru constant; for $d=3$, $\lambda_3^4=\lambda_3^3+1$ gives the fourth. Table~\ref{tab:meru} records the rates and the residual at the measure-matched row. Choosing $a=1/3$ instead gives the common margin in \eqref{eq:maxgap} for both channels.

\begin{table}[t]
\caption{Two compact instances of the soft family. The context count refers to differential outputs.}
\label{tab:meru}
\centering\renewcommand{\arraystretch}{1.14}
\begin{tabular}{@{}lrr@{}}
\toprule
 & $d=2$ & $d=3$\\\midrule
Conformations & 5 & 7\\
$Q$-contexts & 5 & 6\\
All-policy pairs & 10 & 12\\
$\lambda_d$ & 1.465571231877 & 1.380277569098\\
$\rho_d$ (bits/use) & 0.551463089746 & 0.464958417216\\
$a_0=p_d/2$ & 0.158836098086 & 0.137754020500\\
$\delta(a_0)$ (bits) & 0.472681748671 & 0.440572414461\\
\bottomrule
\end{tabular}
\end{table}

The same constrained-entropy rates occur in the noiseless refractory P300 model \cite{MayyaMainsahReeves2016}. There, the output event sequence is itself runlength constrained, giving an entropy upper bound that also applies with feedback. In the present channel the differential output is noisy even on the desired process, and a finite output-context test measure accounts for every premature request. The matching argument and the results below give the common residual \eqref{eq:familyresidual}, the exact endpoint of the capacity plateau, and the trajectory-level converse at $a=1/3$.

\subsection{Capacity of the designed family}

The inverse construction gives the following operational capacity result. Its converse accounts for every request, and its lower bound uses the prescribed zero-error signaling paths.

\begin{theorem}[Soft refractory family]
\label{thm:softfamily}
Let $a_{\rm c}\in(0,1)$ be the nonzero root of $H_2(a)=a$. For every $d\ge2$ and $0\le a\le a_{\rm c}$,
\begin{equation}
 C_{0,\mathrm{fb}}(\mathsf{SRT}_d(a))
 =C_{\mathrm{fb}}(\mathsf{SRT}_d(a))=\rho_d.
 \label{eq:familycapacity}
\end{equation}
For \eqref{eq:refractorytest}, \eqref{eq:normalpotential}, and \eqref{eq:shadowpotential}, every desired normal action and every autonomous action has zero residual. Every premature action in $F_k$ has the same residual
\begin{equation}
 \Delta=\delta(a)=H_2(a)-a.
 \label{eq:familyresidual}
\end{equation}
For $0<a<a_{\rm c}$ this residual is positive. Hence every reliable capacity-approaching sequence has vanishing expected premature-action density and vanishing normalized output divergence from $T$.
\end{theorem}

\begin{IEEEproof}
The desired zero-error scheme gives the lower bound independently of $a$. Lemma~\ref{lem:allpolicy} verifies the all-policy closure and support condition for the converse. The normal-state Bellman equalities follow from the identities and substitutions in Appendix~\ref{app:bellman-equalities}. Equation~\eqref{eq:Jrequired} and the shadow extension \eqref{eq:shadowpotential} give equality on every autonomous row. The preceding design calculation gives residual $-\kappa(a)=\delta(a)$ on every premature request.

The function $H_2(a)-a$ is strictly concave, is zero at $a=0$, is positive at $a=1/2$, and is negative at $a=1$. It has a unique further zero $a_{\rm c}$ and is nonnegative on $[0,a_{\rm c}]$. Theorem~\ref{thm:exactgap} provides the upper bound on that interval, including its endpoints. Theorem~\ref{thm:stability} gives the strict conclusions in its interior.
\end{IEEEproof}

Numerically,
\begin{equation}
 a_{\rm c}=0.772907804781\ldots.
\end{equation}

\subsection{Boundary of the capacity plateau}

At the two endpoints the capacity equality remains, but this residual
no longer forces premature requests to disappear. The next result shows
that the interval is exact. Beyond it, occasional premature requests
increase the achievable information rate.

\begin{theorem}[Exact capacity plateau and zero-error separation]
\label{thm:capacityplateau}
For every $d\ge2$ and $a\in[0,1]$,
\begin{equation}
 C_{\mathrm{fb}}(\mathsf{SRT}_d(a))=\rho_d
 \quad\Longleftrightarrow\quad a\le a_{\rm c}.
 \label{eq:capacityplateau}
\end{equation}
Moreover,
\begin{equation}
 C_{0,\mathrm{fb}}(\mathsf{SRT}_d(a))=\rho_d,
 \quad 0\le a<1.
 \label{eq:zeroerrorfullinterval}
\end{equation}
Thus Shannon and zero-error feedback capacities are different when
$a_{\rm c}<a<1$.

More specifically, retain activation probability $p_d$ at $R$, use a
premature request with probability $\varepsilon$ at $F_1$, and use
ordinary requests at all other recovery states. Independent action
randomization under this policy has an achievable information rate
\begin{equation}
 \mathcal I_a(\varepsilon)
 =\rho_d+\frac{p_d\{a-H_2(a)\}}{1+dp_d}\,\varepsilon
       +O(\varepsilon^2)
 \label{eq:plateauderivative}
\end{equation}
as $\varepsilon\downarrow0$, for each fixed $d$ and $a$.
\end{theorem}

\begin{IEEEproof}
Appendix~\ref{app:plateau} constructs the finite output filter of this
policy and proves \eqref{eq:plateauderivative}. Its premature-request
density is linear in $\varepsilon$, whereas its relative entropy rate
from the original test process is $O(\varepsilon^2)$. The exact gap
identity therefore gives the displayed first-order term. The appendix
also proves operational achievability by resetting the conformation
between finite strategy-tree blocks. Since $a-H_2(a)>0$ for
$a>a_{\rm c}$, a sufficiently small positive $\varepsilon$ gives a rate
strictly above $\rho_d$. Theorem~\ref{thm:softfamily} proves the other
direction of \eqref{eq:capacityplateau}.

For $0<a<1$, the support of every physical transition is independent of
$a$. Fix any finite feedback code, including its encoder and decoder.
The input--output histories having positive probability under that code
are consequently the same for every interior parameter. Whether the
code has zero error is therefore independent of $a$, as in the
support-based characterization of zero-error communication
\cite{Shannon1956}. This applies at every blocklength. Theorem~\ref{thm:softfamily}
at $a=1/2$ gives \eqref{eq:zeroerrorfullinterval} for all $0<a<1$; that
theorem also covers $a=0$.
\end{IEEEproof}

The transition at $a_{\rm c}$ concerns the usefulness of an additional
signaling action. A premature request becomes profitable when the
first-order gain $a-H_2(a)$ is positive; the accompanying change in the
output law has only a second-order divergence cost. The full Shannon
capacity above the plateau is not determined by this perturbation.
At $a=1$ the channel support changes, and
\eqref{eq:zeroerrorfullinterval} makes no zero-error assertion for that
endpoint.

\begin{figure}[t]
\centering
\resizebox{0.98\columnwidth}{!}{%
\begin{tikzpicture}
\begin{axis}[
 width=\columnwidth,height=5.6cm,
 xmin=0,xmax=1,ymin=-1.04,ymax=.76,
 axis lines=left,axis line style={draw=IBInk},
 xlabel={Response probability $a$ at $J$},
 ylabel={Residual $\delta(a)$ [bits]},
 label style={font=\small},tick label style={font=\scriptsize},
 xtick={0,.333333,.7729078,1},
 xticklabels={$0$,$1/3$,$a_{\rm c}$,$1$},
 ytick={-1,-.5,0,.5849625},
 yticklabels={$-1$,$-0.5$,$0$,$\log_2(3/2)$},
 grid=major,grid style={draw=IBGuide,densely dotted},
 clip=false,
 ]
\path[fill=IBShade] (axis cs:0,0) rectangle (axis cs:.772907804781,.72);
\addplot[IBInk,densely dashed,thin] coordinates {(0,0)(1,0)};
\addplot[IBInk,line width=1.05pt] table[x=a,y=gap,col sep=comma]{figures/gap_curve.csv};
\addplot[only marks,mark=*,mark size=2.1pt,IBInk] coordinates {(.333333333333,.584962500721)};
\addplot[only marks,mark=*,mark size=1.7pt,IBInk] coordinates {(.772907804781,0)};
\draw[IBInk,densely dashed] (axis cs:.333333333333,0)--(axis cs:.333333333333,.584962500721);
\node[font=\scriptsize,anchor=south,IBInk] at (axis cs:.52,.64) {largest common gap};
\node[font=\scriptsize,align=center,anchor=center,IBInk] at (axis cs:.38,.18) {strict inverse matching};
\node[font=\scriptsize,align=left,anchor=west,IBInk] at (axis cs:.09,-.68) {$a_{\rm c}=0.7729078\ldots$\\$H_2(a_{\rm c})=a_{\rm c}$};
\end{axis}
\end{tikzpicture}}
\caption{The common premature-action residual $H_2(a)-a$ is independent of the recovery order. Shannon feedback capacity equals $\rho_d$ exactly on $0\le a\le a_{\rm c}$, where the residual is nonnegative. It exceeds $\rho_d$ for $a>a_{\rm c}$, while zero-error feedback capacity remains $\rho_d$ for $a<1$. The maximum residual occurs at $a=1/3$, where the corrected information increment has no output-induced fluctuation conditional on the current state--context pair and request. The plotted residual is an action penalty, not the capacity above the threshold.}
\label{fig:gapcurve}
\end{figure}
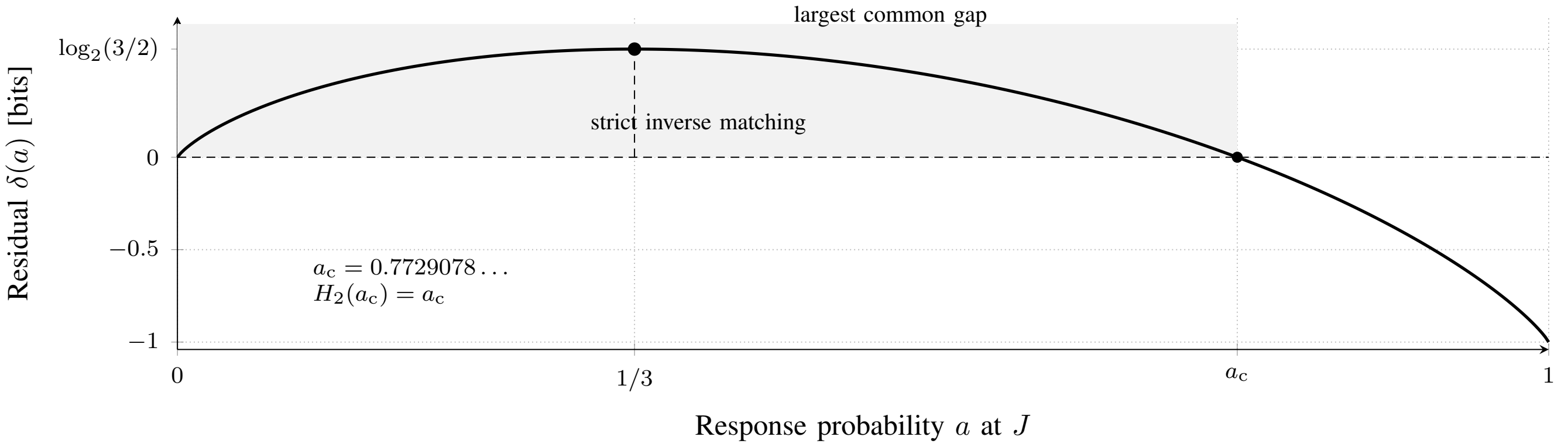

\subsection{Trajectory-wise matching and a finite-block converse}

The Gibbs choice $a=1/3$ has a trajectory-level consequence in addition to maximizing the common residual. The realized dynamic surprise equals its conditional mean for every state--context pair and request, so the Bellman relation holds separately for each possible output. Summing this relation gives an exact likelihood identity from which a finite-block converse follows. In reduced coordinates
write $Z_t=(C_t,Q_t)$ and
\begin{equation}
 N_{\mathrm p}(n)=\sum_{t=1}^n\mathbf1\{C_{t-1}\in\{F_1,\ldots,F_d\},\ U_t=1\}.
 \label{eq:refractory-premature-count}
\end{equation}
Here $W(D^n\Vert U^n,R)$ denotes the causal law of the reduced
channel starting in $R$, and $T_{D^n}$ is its output test law.

\begin{theorem}[Trajectory-wise matching]
\label{thm:trajectorymatching}
Fix $d\ge2$ and $a=1/3$. Write $\rho=\rho_d$,
$\delta_\star=\log_2(3/2)$, and let $v$ be the potential in
\eqref{eq:normalpotential}--\eqref{eq:shadowpotential} at this
parameter. For every causal reduced strategy, every blocklength,
and every trajectory of positive probability,
\begin{align}
 &\log_2\frac{W(D^n\Vert U^n,R)}{T_{D^n}(D^n)}\nonumber\\
 &\quad=n\rho+v(z_0)-v(Z_n)
              -\delta_\star N_{\mathrm p}(n),
 \quad z_0=(R,q_0).
 \label{eq:trajectorymatching}
\end{align}
Let $\Gamma_v=v(z_0)-\min_{z\in\mathcal Z_{\rm all}}v(z)$.
Every length-$n$ feedback code with $M_n$ equiprobable messages
satisfies, for every $k\ge0$,
\begin{equation}
 \Pr\{\widehat M=M,\ N_{\mathrm p}(n)\ge k\}
 \le \frac{2^{n\rho+\Gamma_v-\delta_\star k}}{M_n}.
 \label{eq:correct-violation-bound}
\end{equation}
The assertion permits private encoder randomization.
\end{theorem}

\begin{IEEEproof}
Use the corrected increment $J_t$ from
\eqref{eq:realized-correction}, with $\nu=0$ and with $(U,D,C)$
as the reduced input, output, and physical state. Every
deterministic row has
\begin{equation}
 J_t=\rho-\delta_\star
       \mathbf1\{C_{t-1}=F_j\text{ for some }j,\ U_t=1\},
 \label{eq:pointwise-corrected}
\end{equation}
because its realized increment equals its conditional mean in
\eqref{eq:conditional-surprise}. It remains to check the
stochastic activation and $J$ rows.

For an activation at $(R,q_0)$, output zero gives log likelihood
ratio $\rho-\ell$ and successor potential $\ell$, whereas
output one gives log likelihood ratio $(d+1)\rho$ and successor
potential $-d\rho$. The current potential is zero, so both
corrected increments equal $\rho$. At $(R,q_1)$ the two log
likelihood ratios are $\rho-1$ and $(d+1)\rho-1$; their
successor potentials are $\ell$ and $\ell-d\rho$, respectively.
Subtracting the current potential $\ell-1$ again gives
$\rho$ for either output.

For the row at $(J,q_0)$ and a general $0<a<1$, direct
substitution gives
\begin{equation}
 J_t-\rho=
 \begin{cases}
  \log_2(1-a)-\kappa(a),&D_t=0,\\
  1+\log_2a-\kappa(a),&D_t=1.
 \end{cases}
 \label{eq:J-fluctuations}
\end{equation}
At $a=1/3$, both values are zero, since
$\kappa(1/3)=\log_2(2/3)=1+\log_2(1/3)$.
This proves \eqref{eq:pointwise-corrected} for every row.
Summing the increments telescopes the potential and proves
\eqref{eq:trajectorymatching}.

For the coding bound, let $K$ contain the encoder's private
randomness, chosen independently of the message and channel
noise. The encoder is then a deterministic causal function of
$(M,K,D^{t-1})$. Denote the true law of $(M,K,D^n)$ by
$\mathbb P$ and compare it with
\begin{equation}
 \mathbb Q=P_M P_K T_{D^n}.
\end{equation}
The likelihood ratio on the support of $\mathbb P$ is
$W(D^n\Vert U^n,R)/T_{D^n}(D^n)$.
It is zero off that support. Under $\mathbb Q$, the message
is uniform and independent of $D^n$, and hence
$\mathbb Q\{\widehat M=M\}=1/M_n$.
On the event $N_{\mathrm p}(n)\ge k$,
\eqref{eq:trajectorymatching} bounds this likelihood ratio by
$2^{n\rho+\Gamma_v-\delta_\star k}$. Integrating over the
correct-decoding event gives
\eqref{eq:correct-violation-bound}.
\end{IEEEproof}

Writing $R_n=n^{-1}\log_2M_n$, the case $k=0$ gives
\begin{equation}
 1-P_e^{(n)}\le 2^{\Gamma_v-n(R_n-\rho)}.
 \label{eq:matched-strong-converse}
\end{equation}
Thus the correct-decoding probability decreases exponentially
at any fixed rate above $\rho$. For $\alpha>0$, the same
argument gives
\begin{equation}
 \Pr\{N_{\mathrm p}(n)\ge\alpha n\}
 \le P_e^{(n)}+
       2^{\Gamma_v-n(R_n-\rho+\delta_\star\alpha)}.
 \label{eq:matched-violation-tail}
\end{equation}
At rates tending to capacity, successfully decoded trajectories
with a fixed positive premature-action density are therefore
exponentially rare. No concentration approximation is used in
these finite-block bounds.

For general $0<a<1$, \eqref{eq:J-fluctuations} describes all
fluctuations of the martingale in
\eqref{eq:surprise-martingale}. If
\begin{equation}
 N_J(n)=\sum_{t=1}^n\mathbf1\{C_{t-1}=J\},
 \label{eq:J-count}
\end{equation}
its sum of conditional increment variances is
\begin{equation}
 \langle\mathcal M\rangle_n
 =a(1-a)\left[1+\log_2\frac{a}{1-a}\right]^2N_J(n).
 \label{eq:refractory-quadratic-variation}
\end{equation}
Indeed, the two increments in \eqref{eq:J-fluctuations} have
mean zero and differ by $1+\log_2(a/(1-a))$; all other
martingale increments vanish. Since $J$ is reachable, $a=1/3$ is
the unique interior parameter at which the martingale vanishes
identically for every causal strategy. It also uniquely maximizes
the residual in \eqref{eq:maxgap}. A strategy that never visits $J$
has no such fluctuations at any parameter. For a single receiver context, this is the
general effect of the Gibbs row: \eqref{eq:singlecontext}
makes $\log_2(W_Y(y\mid s,x)/T(y\mid q))+v(Z^+(y))$
constant on its support.

\section{Identification of off-policy dynamics}
\label{sec:interpretation}

The inverse construction gives a family of physical mechanisms for the same optimal signaling behavior. For every $a\in[0,a_{\rm c}]$, the desired input--output law is unchanged and the channel capacity equals $\rho_d$, while the responses to premature requests vary with $a$. Observations confined to the prescribed behavior therefore leave this physical parameter unidentified. For $0<a<a_{\rm c}$, each visit to the parameter-dependent row requires a premature request with positive Bellman residual. The communication deficit consequently bounds the information obtainable from those visits.

We first identify exactly which observations distinguish the mechanisms. We then combine that likelihood calculation with the rate-gap identity and construct sparse probes that permit consistent estimation at asymptotic capacity.

\begin{proposition}[On-policy nonidentifiability]
\label{prop:identification}
Fix $d\ge2$ and one causal reduced policy that never makes a
premature request. Suppose its rules and initial randomization
do not depend on $a$. For every blocklength, its joint
input--output law is identical for all $a\in[0,1]$.
An experiment that enters $J$ and records its next differential
output instead obtains one $\Bern(a)$ observation. Repeated
experiments starting from known $R$ give independent
observations.
\end{proposition}

\begin{IEEEproof}
Starting at $R$, a policy without premature requests never
enters $S_k$ or $J$. Every visited channel row is independent
of $a$, so induction in time proves equality of the laws.
To enter $J$, activate once, use ordinary recovery requests
until $F_1$, and then use $U=1$. The next conformation is $J$
with certainty, and its next output is $\Bern(a)$. After
output zero the state is $R$; after output one, $d$ ordinary
recovery requests return to $R$. Each repetition uses fresh
channel noise from the same known state, proving independence.
\end{IEEEproof}

A trial that stops upon this return to $R$ uses
$d+2+ad$ channel uses on average. The sample mean of $m$ such observations
has variance $a(1-a)/m$. The following result applies
also when the experiment is adaptive and is combined with
communication. We use the visit count $N_J(n)$ from
\eqref{eq:J-count} and the premature-request count
$N_{\mathrm p}(n)$ from \eqref{eq:refractory-premature-count}.

\subsection{Information acquired by an adaptive procedure}

Fix a horizon $n$. Let $M$ be a message, possibly a constant
for a pure identification experiment, and let $K$ contain
the procedure's private randomization. Its requests have the
form
\begin{equation}
 U_t=\phi_t(M,K,D^{t-1}),\quad 1\le t\le n.
 \label{eq:identification-procedure}
\end{equation}
The functions $\phi_t$ and the joint law of $(M,K)$ are the
same under every parameter being compared; channel noise is
independent of this initial randomization. Thus the procedure
can adapt to its observations, but is not told the true
parameter. Define the full transcript
\begin{equation}
 \mathsf H_n=(M,K,U^n,D^n)
 \label{eq:identification-transcript}
\end{equation}
and write $P_a^{\mathsf H_n}$ for its law at parameter $a$.
The inputs and known initial state determine each conformation
from the past outputs. An observer given less information,
such as only $(U^n,D^n)$ or $D^n$, sees a function of this
full transcript.

\begin{theorem}[Adaptive identification information]
\label{thm:adaptive-identification}
For $0<a,a'<1$ and any procedure
\eqref{eq:identification-procedure},
\begin{align}
 D(P_a^{\mathsf H_n}\|P_{a'}^{\mathsf H_n})
 &=\mathbb E_a N_J(n)\,D_{\rm B}(a\|a')\nonumber\\
 &\le\mathbb E_a N_{\mathrm p}(n)\,D_{\rm B}(a\|a'),
 \label{eq:adaptive-KL}
\end{align}
where $D_{\rm B}(a\|a')=D(\Bern(a)\|\Bern(a'))$.
For every parameter-independent observation obtained from the
full transcript, its relative entropy is no larger than the
quantity in \eqref{eq:adaptive-KL}.
\end{theorem}

\begin{IEEEproof}
For interior parameters the channel transition supports agree.
The initial law of $(M,K)$ is also the same under the two
parameters. Conditional on $(M,K,D^{t-1})$, the request and
conformation are known. Their causal likelihood factors
therefore cancel in the ratio, and every channel row except
the row at $J$ is independent of the parameter. Consequently,
on the common support,
\begin{align}
 &\log_2\frac{dP_a^{\mathsf H_n}}{dP_{a'}^{\mathsf H_n}}\nonumber\\
 &\quad=\sum_{t:C_{t-1}=J}
 \left\{D_t\log_2\frac{a}{a'}
 +(1-D_t)\log_2\frac{1-a}{1-a'}\right\}.
 \label{eq:adaptive-log-likelihood}
\end{align}
The event $C_{t-1}=J$ is determined before $D_t$ is observed.
On that event, $D_t$ has conditional law $\Bern(a)$ under
$P_a$. Conditional expectation of each summand is thus
$D_{\rm B}(a\|a')$, which proves the equality in
\eqref{eq:adaptive-KL}.

Starting from $R$, each use at $J$ is preceded by a distinct
premature request. That request initiates a shadow countdown
which ends with this use at $J$; it cannot generate a second
use at $J$ without another premature request. Some countdowns
may be unfinished at time $n$. Hence
$N_J(n)\le N_{\mathrm p}(n)$ on every trajectory, proving the
inequality. Applying relative-entropy data processing to the
observation map proves the last assertion. An independently
randomized observation can be handled by adjoining its seed
to $K$.
\end{IEEEproof}

The same likelihood calculation gives the Fisher information
of the full transcript. With natural logarithms in the score,
\begin{equation}
 \mathcal J_n(a)=\frac{\mathbb E_aN_J(n)}{a(1-a)}.
 \label{eq:adaptive-Fisher}
\end{equation}
Indeed, the score is
$\sum_{t:C_{t-1}=J}(D_t-a)/(a(1-a))$.
Its summands have conditional mean zero, so cross terms vanish
when its square is averaged, and each use at $J$ contributes
$1/(a(1-a))$. Equation~\eqref{eq:adaptive-Fisher} is an
information identity; it does not assume any particular
estimator.

\subsection{Communication rate and discrimination}

Suppose the same procedure sends $M_n$ equiprobable messages
with decoding error probability $P_{e,a}^{(n)}$ when the true
parameter is $a$. For $0<a<a_{\rm c}$, let $v_a$ be the
potential in \eqref{eq:normalpotential}--\eqref{eq:shadowpotential}
and define
\begin{equation}
 G_n(a)=n\rho_d+\Span(v_a)+1
           -(1-P_{e,a}^{(n)})\log_2M_n.
 \label{eq:identification-rate-deficit}
\end{equation}
The finite-block converse makes $G_n(a)$ nonnegative. For a reliable code sequence with rates tending to $R$, $G_n(a)/n$ tends to the communication deficit $\rho_d-R$.

\begin{corollary}[Rate--discrimination bound]
\label{cor:rate-identification}
Fix $0<a<a_{\rm c}$ and $0<a'<1$ with $a'\ne a$.
Every parameter-independent communication and identification
procedure satisfies
\begin{equation}
 D(P_a^{\mathsf H_n}\|P_{a'}^{\mathsf H_n})
 \le \frac{D_{\rm B}(a\|a')}{\delta(a)}G_n(a).
 \label{eq:rate-KL-bound}
\end{equation}
Let a test decide between $a$ and $a'$ using the full
transcript or any parameter-independent observation of it.
Write
\begin{align}
 \alpha_n&=\Pr_a\{\text{test decides }a'\},\nonumber\\
 \beta_n&=\Pr_{a'}\{\text{test decides }a\}.
 \label{eq:test-errors}
\end{align}
Then
\begin{align}
 &(1-\alpha_n)\log_2\frac1{\beta_n}\nonumber\\
 &\quad\le
 \frac{D_{\rm B}(a\|a')}{\delta(a)}G_n(a)
       +H_2(\alpha_n).
 \label{eq:rate-test-bound}
\end{align}
In particular, if $P_{e,a}^{(n)}\to0$,
$n^{-1}\log_2M_n\to R$, and $\alpha_n\to0$, then
\begin{equation}
 \limsup_{n\to\infty}\frac1n\log_2\frac1{\beta_n}
 \le\frac{D_{\rm B}(a\|a')}{\delta(a)}(\rho_d-R).
 \label{eq:rate-test-exponent}
\end{equation}
\end{corollary}

\begin{IEEEproof}
Drop the nonnegative output-divergence term from
\eqref{eq:codegap} and use the common premature-action residual
$\delta(a)>0$. The result is
$\delta(a)\mathbb E_aN_{\mathrm p}(n)\le G_n(a)$.
Combining this inequality with \eqref{eq:adaptive-KL} proves
\eqref{eq:rate-KL-bound}.

Data processing from the full transcript to the binary test
decision gives
\begin{equation}
 D_{\rm B}(1-\alpha_n\|\beta_n)
 \le D(P_a^{\mathsf H_n}\|P_{a'}^{\mathsf H_n}).
\end{equation}
Expanding the binary divergence shows
\begin{align}
 D_{\rm B}(1-\alpha_n\|\beta_n)
 &=(1-\alpha_n)\log_2\frac1{\beta_n}
       -H_2(\alpha_n)\nonumber\\
 &\quad-\alpha_n\log_2(1-\beta_n)\nonumber\\
 &\ge(1-\alpha_n)\log_2\frac1{\beta_n}
       -H_2(\alpha_n).
\end{align}
This proves \eqref{eq:rate-test-bound}, with the usual
extended-value conventions at zero probabilities. Divide by
$n$ and apply the stated limits to obtain
\eqref{eq:rate-test-exponent}.
\end{IEEEproof}

Thus a reliable capacity-approaching code has zero
discrimination exponent per channel use whenever
$\limsup_n\alpha_n<1$. This does not prevent consistent
identification: the number of informative interventions can
tend to infinity while their density tends to zero. The next
construction realizes both limits with a fixed blocklength
and a parameter-independent schedule.

\subsection{Consistent identification at capacity}

\begin{proposition}[Sparse probing]
\label{prop:sparse-probing}
Let $\tau=2d+2$ and choose integers $m_n$ such that
$m_n\to\infty$ and $m_n=o(n)$. For all sufficiently large $n$, there are
parameter-independent length-$n$ codes and estimators
$\widehat a_n$ with zero decoding error for every
$a\in[0,1]$, rate tending to $\rho_d$, and
\begin{equation}
 \mathbb E_a(\widehat a_n-a)^2
 =\frac{a(1-a)}{m_n}.
 \label{eq:sparse-probing-MSE}
\end{equation}
They are therefore capacity-approaching for every
$a\in[0,a_{\rm c}]$ while consistently estimating $a$.
\end{proposition}

\begin{IEEEproof}
Starting from $R$, use the following fixed request sequence
as one probe block:
\begin{equation}
 1\,0^{d-1}\,1\,0^{d+1}.
 \label{eq:fixed-probe-block}
\end{equation}
Its length is $\tau$. The first request enters $F_d$;
the next $d-1$ requests reach $F_1$; and the second
request equal to one enters $J$. The output at position
$d+2$ is consequently $\Bern(a)$. The final $d$ zero
requests return the conformation to $R$ for either value
of this output. Thus each block begins and ends in known
$R$, and $m_n$ repetitions give independent samples
$\Xi_1,\ldots,\Xi_{m_n}$ with law $\Bern(a)$. Their positions
are known to both encoder and receiver.

After these probes, send a distinct legal $(d,\infty)$
request string of length
$\ell_n=n-\tau m_n-d$, and append $d$ zero requests.
For all sufficiently large $n$, $\ell_n$ is positive.
The finite-delay observer of Lemma~\ref{lem:observer}
recovers this string with zero error. This part of the
code never enters the shadow states, so its decoding
property holds for every value of $a$. There are
$M_n=N_d(\ell_n)$ possible message strings, and
\eqref{eq:countbounds} gives
\begin{equation}
 \log_2M_n=\ell_n\rho_d+O(1).
 \label{eq:sparse-probing-rate}
\end{equation}
The constant depends on $d$, not on $n$ or $a$.
Since $m_n=o(n)$, the rate tends to $\rho_d$.
Finally, take
$\widehat a_n=m_n^{-1}\sum_{j=1}^{m_n}\Xi_j$.
Independence of the probe samples gives
\eqref{eq:sparse-probing-MSE}, which tends to zero.
\end{IEEEproof}

The same construction gives a quantitative achievable
discrimination tradeoff. For $0<a<a'<1$, choose any
$t\in(a,a')$ and decide $a'$ when $\widehat a_n\ge t$.
The two testing errors obey
\begin{equation}
 \alpha_n\le2^{-m_nD_{\rm B}(t\|a)},\quad
 \beta_n\le2^{-m_nD_{\rm B}(t\|a')}.
 \label{eq:probe-test-errors}
\end{equation}
For example, an exponential Markov inequality gives
\begin{align}
 \Pr_a\{\widehat a_n\ge t\}
 &\le\inf_{s\ge0}
       (1-a+a2^s)^{m_n}2^{-s m_n t}\nonumber\\
 &=2^{-m_nD_{\rm B}(t\|a)};
\end{align}
the lower-tail calculation under $a'$ is the same with
$s\le0$. Taking $m_n=\lfloor n^\gamma\rfloor$ for any
$0<\gamma<1$ makes both errors vanish while preserving
the rate limit $\rho_d$, which equals capacity on
$[0,a_{\rm c}]$. Alternatively, if
$\tau m_n/n\to\eta\in(0,1)$, the achieved communication
rate is $(1-\eta)\rho_d$ and the two error exponents per
channel use are at least
$\eta D_{\rm B}(t\|a)/\tau$ and
$\eta D_{\rm B}(t\|a')/\tau$, respectively.
The converse and construction quantify how positive
discrimination exponents require a nonzero communication
rate deficit within the strict-residual regime.

\section{Conclusion}

The inverse problem studied here, inspired by theoretical biology, fixes a signaling behavior and seeks physical channel dynamics under which that behavior supports capacity-achieving codes. Dynamic measure matching gives sufficient conditions, in terms of dynamic surprise, that apply to every causal strategy on the resulting channel. The continuation potential accounts for signaling opportunities enabled or removed by a present action, and the exact dual slack separates action loss from mismatch of the output process. For fixed architecture, test law, and potential, channel-row synthesis is a convex problem under convex physical restrictions. The minimax row construction enforces the requirement that one physical law work for every compatible receiver context.

Graph and refractory constructions establish capacity attainment through explicit zero-error codes. A finite irreducible constrained graph of positive entropy admits a soft realization whose violation penalty is the entropy of an autonomous path. The noisy refractory family (that models biochemistry) determines the exact interval over which the prescribed behavior remains optimal and exhibits a separation between Shannon and zero-error feedback capacities over an interval beyond it. In this family, the Gibbs choice maximizes the premature-action penalty and makes dynamic surprise equal its conditional mean on every possible transition. Its trajectory-wise likelihood identity gives a finite-block bound on successful decoding with premature requests and an exponential strong converse. More generally, a uniform positive residual on specified departures implies that every reliable sequence of codes approaching capacity uses those departures with vanishing expected frequency.

Observations confined to the prescribed behavior can leave the mechanism unidentified. In the refractory family, learning the unused parameter-dependent dynamics requires visits initiated by premature requests. Their residual bounds the discrimination information acquired by an adaptive experiment that also communicates. Positive discrimination exponents require a positive rate loss within the strict plateau, while sparse probes permit consistent estimation at asymptotic capacity. An optimization-based scientific explanation therefore also needs independent restrictions on the mechanism and resource law. Further inverse-design questions concern realizability under those restrictions, the joint choice of architecture and matching quantities, and the number of states, delay, and measurement accuracy needed to obtain a strict behavioral penalty.

\section*{Acknowledgment}
A conversation with Oron Sabag motivated this inverse approach.

\section*{AI use}
ChatGPT 5.6-Sol and then ChatGPT 6 were used to support the mathematical development, writing, and drawing figures.

\begingroup
\newcommand{\BIBdecl}{\sloppy\emergencystretch=3em}
\sloppy
\emergencystretch=2em
\bibliographystyle{IEEEtran}
\bibliography{references}
\endgroup

\appendices

\section{Predictive surprise and the continuation potential}
\label{app:surprise-interpretation}

The divergence has a Bayesian interpretation under a matched predictive law. Let $K=(S,X)$ have prior $\mu_q$ and suppose
$T(y\mid q)=\sum_k\mu_q(k)W_Y(y\mid k)$. For $\mu_q(k)>0$,
\begin{equation}
 \log\frac{W_Y(y\mid k)}{T(y\mid q)}
  =\log\frac{\mu_q(k\mid y)}{\mu_q(k)}.
 \label{eq:bayesratio}
\end{equation}
Consequently, $r_T(k,q)$ is the expected log posterior-to-prior update conditional on the actual hypothesis $k$. Averaging over $k$ gives the expected posterior divergence, or Bayesian surprise \cite{IttiBaldi2005,Varshney2013Surprise,Varshney2019Creativity}:
\begin{align}
 \sum_k\mu_q(k)r_T(k,q)
 &=I(K;Y\mid Q=q)\nonumber\\
 &=\mathbb E_Y D(\mu_q(\cdot\mid Y)\|\mu_q).
 \label{eq:bayesaverage}
\end{align}
Off policy, the same $T$ remains a reference law for the converse; it need not be the actual output-predictive distribution.

When a policy makes the Bellman equation tight, $v$ is its differential value for the reward $r_T-\nu b$. For example, let $\tau$ be the hitting time of a reference state $z_\circ$ under that policy, and suppose $\mathbb E_z\tau<\infty$. Then
\begin{equation}
 v(z)-v(z_\circ)
 =\mathbb E_z\sum_{t=1}^{\tau}
       \{r_T(Z_{t-1},X_t)-\nu b(S_{t-1},X_t)-\rho\}.
 \label{eq:surprise-to-go}
\end{equation}
This follows by stopping the tight Bellman recursion; bounded rewards and $\mathbb E\tau<\infty$ justify the limit \cite{Puterman1994}.

Recursive information-to-go was developed by Tishby and Polani \cite{TishbyPolani2011}. Nonmyopic Bayesian information gain \cite{SunGomezSchmidhuber2011,HuanMarzouk2016} and differential surprise-to-go in average-surprise control \cite{SenneshEtAl2023} also predate the present construction. Here the quantity emerges from a coding theorem: it enters an operational capacity converse and is used to design the channel. Its important feature is temporal: two actions may have the same immediate divergence and different continuation values. The refractory example makes this difference explicit.

\section{Constrained-system enumeration and effective-action realizations}
\label{app:supporting}

This appendix gives state-conditioned path enumeration and an effective-action realization criterion for the constrained systems of Section~\ref{sec:constraints}.

\subsection{Goulden--Jackson enumeration}
\label{app:enumeration}

The Goulden--Jackson cluster method obtains rational generating functions from overlaps of forbidden words \cite{GouldenJackson1979}. Applications to constrained coding include periodic finite-type shifts \cite{BealCrochemoreMoisionSiegel2011} and skip-sliding-window codes, whose good-word cluster formulation was related directly to finite-state enumeration \cite{WuTandonVarshneyMotani2021}. A systematic finite-type treatment relates the convergence radius to the constrained graph's spectral radius \cite{ShenShangguanLinGe2025}.

For state-conditioned continuation counts,
\begin{equation}
 \mathbf F(z)=\sum_{n\ge0}A^n\mathbf1 z^n=(I-zA)^{-1}\mathbf1.
 \label{eq:GF}
\end{equation}
Let $\ell$ be a positive left eigenvector satisfying $\ell^{\mathsf T}A=\lambda\ell^{\mathsf T}$. With $R=\lambda^{-1}$, the simple Perron pole gives
\begin{equation}
 \lim_{z\uparrow R}(1-z/R)\mathbf F(z)
   =\frac{\ell^{\mathsf T}\mathbf1}{\ell^{\mathsf T}r}\,r.
 \label{eq:residue}
\end{equation}
Indeed, the resolvent equals $(1-z/R)^{-1}r\ell^{\mathsf T}/(\ell^{\mathsf T}r)+O(1)$ near $R$. Consequently, scalar enumeration gives $\rho=-\log_2R$, and state- or suffix-refined residues give $h$ up to a constant. The scalar growth constant alone does not determine the potential.

For the $(d,\infty)$ constraint, successive ones are separated by at least $d$ zeros, so the forbidden words are $10^k1$, $0\le k<d$. Let $N_d(n)$ count its length-$n$ binary words and $F_d(z)=\sum_{n\ge0}N_d(n)z^n$. The forbidden words have only the symbol $1$ as a proper interword overlap. Writing $C_k$ for the signed cluster series ending in $10^k1$ and $C=\sum_kC_k$, the cluster equations are
\begin{equation}
 C_k=-z^{k+2}-z^{k+1}C,
 \quad F_d(z)=\frac1{1-2z-C}.
\end{equation}
Solving gives
\begin{equation}
 F_d(z)=\frac{1+z+\cdots+z^d}{1-z-z^{d+1}}.
 \label{eq:RLLGF}
\end{equation}
Thus $N_d(n)=N_d(n-1)+N_d(n-d-1)$ for $n\ge d+1$, with $N_d(n)=n+1$ for $0\le n\le d$. The second- and fourth-Meru recurrences correspond to $d=2,3$ \cite{Varshney2008Meru}. In a grouped skip-sliding-window presentation, one graph step represents a group of symbols; entropy must be divided by that group length to obtain rate per physical symbol.

\subsection{Hard effective-action realizations}
\label{app:effective-actions}

A structural route to a converse is to make every physical strategy project onto a legal graph path. The following formulation specifies both simulation and decoding.

\begin{proposition}[Effective-action realization]
\label{prop:quotient}
Suppose a unifilar channel with a known initial state has a map $\eta:\mathcal S\to V$ and a causal edge $E_t=\psi(S_{t-1},X_t)$ such that every physical transition satisfies
\begin{equation}
 \tail(E_t)=\eta(S_{t-1}),\quad
 \eta(S_t)=\head(E_t).
 \label{eq:homomorphism}
\end{equation}
Assume $M-(E^t,Y^{t-1})-Y_t$ is a Markov chain for every feedback code. Then $C_{\mathrm{fb}}\le\log_2\lambda$.

For equality, suppose further that for every compatible $s$ and outgoing edge $e$ there is a selector $x(s,e)$ implementing $e$ for all outputs in its support, and that a decoder recovers the first $n$ edges from $Y^{n+L}$ for every legal path and every channel realization, with fixed $L$. Then
\begin{equation}
 C_{0,\mathrm{fb}}=C_{\mathrm{fb}}=\log_2\lambda.
\end{equation}
\end{proposition}
\begin{IEEEproof}
Conditional data processing and directed-information conservation give
\begin{equation}
 I(M;Y^n)\le I(E^n\!\to Y^n)
             \le I(E^n;Y^n)\le H(E^n).
\end{equation}
More precisely, since every effective path starts at $\eta(s_0)$,
\begin{equation}
 I(M;Y^n)\le\log_2 N_{\eta(s_0)}(n).
 \label{eq:quotient-exact-count}
\end{equation}
Thus exact cluster enumeration also gives a finite-block converse, including its boundary terms. Proposition~\ref{prop:perron} gives the asymptotic upper bound. For achievability, select a length-$n$ graph path as the message. The encoder knows $S_{t-1}$ and implements its next edge with $x(S_{t-1},e_t)$. Equation~\eqref{eq:homomorphism} preserves compatibility after every possible output. Append any legal $L$-edge continuation. The observer recovers the message path without error. Equation~\eqref{eq:countbounds} gives rate $\log_2\lambda$.
\end{IEEEproof}

The homomorphism places every effective action sequence in the legal
path set. Together with the conditional-independence assumption, this
gives the information bound for every physical strategy. The selectors
and observer establish the separate coding lower bound. In the soft
construction of Section~\ref{sec:constraints}, the autonomous-path
entropy additionally quantifies the loss caused by an initiating
undesired action.

\section{Physical representation and Bellman verification}
\label{app:bellman}

This appendix gives the physical coordinate representation, the complete all-policy transition check, and the row-by-row Bellman substitutions for Section~\ref{sec:refractory}.

\subsection{Physical representation and causal equivalence}
\label{app:physical-coordinates}

A two-species label representation retains a bit $B$ and has physical input and readout $X,Y\in\{0,1\}$. Define $U=X\oplus B$. At $(B,R)$, $U=0$ gives $Y=B$ and preserves $(B,R)$; $U=1$ gives a fair binary $Y$, retains $B^+=1-Y$, and enters $F_d$. In every normal or shadow recovery state, $Y=B$ and $B^+=B$; the conformation update is the one in \eqref{tab:channel}. At $J$, with probability $1-a$ the readout is $Y=B$ and the conformation becomes $R$; with probability $a$, $Y=1-B$ and the conformation becomes $F_d$. In both cases $B$ is retained. This defines the full physical kernel without interpreting a repeated readout as emission of the same molecule.

\begin{IEEEproof}[Proof of Lemma~\ref{lem:coordinates}]
On every physical trajectory,
\begin{equation}
 B_t=Y_t\oplus\mathbf1\{C_t=F_d\}.
 \label{eq:species-invariant}
\end{equation}
It holds initially. An activation in state $R$, or an output $D_t=1$ from state $J$, gives $C_t=F_d$ and $B_t=1-Y_t$. Every other transition satisfies $B_t=Y_t$ and $C_t\ne F_d$. Thus the invariant is preserved. It implies that a transition from $F_d$ has differential output one, all other recovery transitions have differential output zero, and a ready activation has a fair differential output. The $J$ row is exactly \eqref{tab:channel}.

Fix the message and any private encoder seed. From $D^{t-1}$ and the known $Y_0$, reconstruct $Y^{t-1}$; the physical encoder then determines its past inputs and unifilar state. This determines $U_t$. Conversely, given a reduced encoder, reconstruct $C_{t-1}$ from $(U^{t-1},D^{t-1})$, reconstruct $Y_{t-1}$, obtain $B_{t-1}$ from \eqref{eq:species-invariant}, and choose $X_t=B_{t-1}\oplus U_t$. These causal maps are inverse. The output maps are bijective, so error probabilities and message sets are unchanged.

Conditioned on $Y^{t-1}$, the input histories $X^t$ and $U^t$ are bijective, and $Y_t\leftrightarrow D_t$ is a bijection. Since $D^{t-1}\leftrightarrow Y^{t-1}$, each conditional mutual-information term in the two directed informations agrees.
\end{IEEEproof}

For the original physical readout $Y$, an equivalent context is $\widetilde Q=(Q,Y_{t-1})$ with
\begin{align}
 \widetilde g((q,y_-),y)&=(g(q,y\oplus y_-),y),\nonumber\\
 \widetilde T(y\mid q,y_-)&=T(y\oplus y_-\mid q).
 \label{eq:physicaltest}
\end{align}
The physical potential is $v(C,Q)$, independent of the species label. Lemma~\ref{lem:coordinates} and \eqref{eq:physicaltest} allow the reduced converse to be applied to the physical channel using context $\widetilde Q$ and potential $v(C,Q)$.

\subsection{All-policy closure and domination}
\label{app:allpolicy}

\begin{table}[t]
\caption{Complete all-policy product-state transitions. Here $j\in\{1,2,3\}$, $1\le k<d$, $F_0=R$, $S_0=J$, and $u_0=q_0$. Entries $U/D$ give the request and output; $*$ denotes either request. At $R$, outputs for $U=1$ are equiprobable; at $J$, their probabilities are $1-a,a$.}
\label{tab:producttransitions}
\centering
\begingroup
\small
\setlength{\tabcolsep}{4pt}
\renewcommand{\arraystretch}{1.14}
\begin{tabular*}{\columnwidth}{@{\extracolsep{\fill}}lcl@{}}
\toprule
Current pair & $U/D$ & Successor pair\\
\midrule
$(R,q_0)$ & $0/0$ & $(R,q_1)$\\
 & $1/0$ & $(F_d,q_1)$\\
 & $1/1$ & $(F_d,q_2)$\\[1pt]
$(R,q_1)$ & $0/0$ & $(R,q_1)$\\
 & $1/0$ & $(F_d,q_1)$\\
 & $1/1$ & $(F_d,q_3)$\\[1pt]
$(F_d,q_j)$ & $0/1$ & $(F_{d-1},g(q_j,1))$\\
 & $1/1$ & $(S_{d-1},g(q_j,1))$\\[1pt]
$(F_{d-1},q_3)$ & $0/0$ & $(F_{d-2},u_{d-2})$\\
 & $1/0$ & $(S_{d-2},u_{d-2})$\\[1pt]
$(F_k,u_k)$ & $0/0$ & $(F_{k-1},u_{k-1})$\\
 & $1/0$ & $(S_{k-1},u_{k-1})$\\[1pt]
$(S_{d-1},q_3)$ & $*/0$ & $(S_{d-2},u_{d-2})$\\
$(S_k,u_k)$ & $*/0$ & $(S_{k-1},u_{k-1})$\\[1pt]
$(J,q_0)$ & $*/0$ & $(R,q_1)$\\
 & $*/1$ & $(F_d,q_2)$\\
\bottomrule
\end{tabular*}
\endgroup
\end{table}

\begin{IEEEproof}[Proof of Lemma~\ref{lem:allpolicy}]
Table~\ref{tab:producttransitions} lists every successor. It proves closure, including after a premature request. Conversely, the desired requests reach both ready contexts, the three $F_d$ contexts, $(F_{d-1},q_3)$, and all known recovery pairs with positive probability. Replacing one recovery request by a premature request reaches the corresponding shadow successor. Countdown reaches every listed shadow pair and $(J,q_0)$. This argument does not use either branch probability at $J$, so it also covers $a=0,1$.

The test laws at $q_0,q_1,q_3$ have full binary support. Every compatible row at $q_2$ emits only one, and every compatible row at $u_k$ emits only zero. Thus every row is dominated, and no completion edge is traversed.
\end{IEEEproof}

\subsection{Bellman equalities}
\label{app:bellman-equalities}

Fix $d\ge2$. Write $\lambda=\lambda_d$, $\rho=\log_2\lambda$, $p=\lambda^{-(d+1)}$, $a_0=p/2$, and $\ell=\log_2(\lambda+1)$. The root equation gives
\begin{align}
 1-p&=\lambda^{-1},&1-a_0&=\frac{\lambda+1}{2\lambda},\label{eq:app-rootid}\\
 -\log_2p&=(d+1)\rho,&-\log_2(1-p)&=\rho,\label{eq:elementaryids}\\
 -\log_2(1-a_0)&=\rho-\ell+1.\label{eq:app-logid}
\end{align}
For the random activation row,
\begin{align}
 D_{\rm B}(1/2\|p)&=\frac{d+2}{2}\rho-1,\label{eq:app-divp}\\
 D_{\rm B}(1/2\|a_0)&=\frac{d+2}{2}\rho-\frac{\ell}{2}.\label{eq:app-diva}
\end{align}
These identities use logarithms to base two. They apply to the test process regardless of the physical probability $a$ at $J$.

Write $v^+=v(C^+,Q^+)$ for the successor potential. At $(R,q_0)$, request zero has divergence $-\log_2(1-a_0)$ and successor value $\ell-1$, so
\begin{equation}
 r_T+\mathbb E v^+=(\rho-\ell+1)+(\ell-1)=\rho.
\end{equation}
Request one has two equiprobable successor values $\ell$ and $-d\rho$. Thus
\begin{equation}
 r_T+\mathbb E v^+
 =\left(\frac{d+2}{2}\rho-\frac\ell2\right)
    +\frac{\ell-d\rho}{2}=\rho.
\end{equation}
At $(R,q_1)$, request zero gives $\rho+\ell-1$. Request one has successor values $\ell$ and $\ell-d\rho$, giving
\begin{equation}
 r_T+\mathbb E v^+
 =\left(\frac{d+2}{2}\rho-1\right)
   +\ell-\frac{d\rho}{2}=\rho+\ell-1.
\end{equation}
Both requests are therefore tight at both ready-state contexts.

The desired recovery requests are deterministic. Their divergence and continuation sums are
\begin{align}
 (F_d,q_1):\quad &(d+1)\rho+(\ell-d\rho)=\rho+\ell,\\
 (F_d,q_2):\quad &0-(d-1)\rho=\rho-d\rho,\\
 (F_d,q_3):\quad &\ell-(d-1)\rho=\rho+\ell-d\rho,\\
 (F_{d-1},q_3):\quad &(\ell-\rho)-(d-2)\rho=\rho+\ell-d\rho,\\
 (F_k,u_k):\quad &0-(k-1)\rho=\rho-k\rho.
\end{align}
The conventions $F_0=R$ and $u_0=q_0$ include $d=2$ and $k=1$. Each expression equals $\rho+v$ at the indicated starting pair.

Let $\kappa(a)=a-H_2(a)$. At $(J,q_0)$ the divergence is $D_{\rm B}(a\|a_0)$, and direct expansion gives
\begin{align}
 &D_{\rm B}(a\|a_0)+(1-a)(\ell-1)-ad\rho\nonumber\\
 &=-H_2(a)+a\{(d+1)\rho+1\}\nonumber\\
 &\quad +(1-a)(\rho-\ell+1)+(1-a)(\ell-1)-ad\rho\nonumber\\
 &=\rho+a-H_2(a)=\rho+\kappa(a).
 \label{eq:app-J}
\end{align}
This calculation remains valid at $a=0,1$ with $0\log 0=0$. The shadow rows have deterministic output zero. At $(S_{d-1},q_3)$,
\begin{align}
 r_T+v^+
 &=(\ell-\rho)-(d-2)\rho+\kappa(a)\nonumber\\
 &=\rho+\ell-d\rho+\kappa(a).
\end{align}
This yields the required equality. At $(S_k,u_k)$,
\begin{equation}
 r_T+v^+=-(k-1)\rho+\kappa(a)=\rho-k\rho+\kappa(a).
\end{equation}
Finally, a premature request has the same immediate output and next $Q$-node as the desired request at the same normal recovery state. Its shadow successor value differs by exactly $\kappa(a)$. Its residual is therefore $-\kappa(a)=H_2(a)-a$. This verifies every row in the all-policy transition table.

\section{A perturbation beyond the capacity plateau}
\label{app:plateau}

Fix $d\ge2$ and $a\in[0,1]$, and abbreviate
$p=p_d$, $\rho=\rho_d$, and $\lambda=\lambda_d$.
For $0\le\varepsilon<1$, consider the stationary causal policy
\begin{equation}
 P(U_t=1\mid C_{t-1})=
 \begin{cases}
 p,&C_{t-1}=R,\\
 \varepsilon,&C_{t-1}=F_1,\\
 0,&\text{otherwise}.
 \end{cases}
 \label{eq:plateaupolicy}
\end{equation}
Each request uses fresh randomization with the displayed probabilities.
The encoder knows the current conformation by unifilarity.
The conformation remains in $\{R,F_d,\ldots,F_1,J\}$ from the initial $R$.

\subsection{Explicit finite output filter}

Every return of the output context to $q_0$ occurs immediately after a
use of $F_1$. That use emits zero under either request, while the fresh
input randomization selects successor $R$ or $J$. Consequently, at every
such return,
\begin{equation}
 P(C_t=R\mid D^t)=1-\varepsilon,\quad
 P(C_t=J\mid D^t)=\varepsilon.
 \label{eq:plateauresetbelief}
\end{equation}
The initial pure-$R$ belief is the sole exception. The first return has
finite expected time: the waiting time for an activation is geometric
with parameter $p$, followed by $d$ recovery uses.

At a return to $q_0$, let $k_\varepsilon$ denote the probability of output one. After output zero, write $(A_\varepsilon,B_\varepsilon)$ for the posterior on $(R,F_d)$. Let $s_\varepsilon$ be the probability of output one from this posterior and $c_\varepsilon$ the resulting posterior probability of $F_d$. Then
\begin{align}
 k_\varepsilon&=(1-\varepsilon)p/2+\varepsilon a,\nonumber\\
 A_\varepsilon
 &=\frac{(1-\varepsilon)(1-p)+\varepsilon(1-a)}{1-k_\varepsilon},\nonumber\\
 B_\varepsilon&=\frac{(1-\varepsilon)p}{2(1-k_\varepsilon)},
 \label{eq:plateaubeliefs}\\
 s_\varepsilon&=A_\varepsilon p/2+B_\varepsilon,\nonumber\\
 c_\varepsilon&=\frac{A_\varepsilon p}{2s_\varepsilon},
 \quad c_0=\frac1{\lambda+1}.
 \label{eq:plateauprobabilities}
\end{align}
Here $A_\varepsilon+B_\varepsilon=1$. Split $q_1$ into $q_1^{\mathrm f}$, for the first zero after $q_0$ with belief $(A_\varepsilon,B_\varepsilon)$, and $q_1^{\mathrm r}$, for subsequent zeros. Another zero
can only have come from $R$, and Bayes' rule gives
\begin{equation}
 q_1^{\mathrm r}:\quad
 \left(P(R),P(F_d)\right)
 =\left(\frac{1-p}{1-p/2},\frac{p/2}{1-p/2}\right).
 \label{eq:plateaurepeatedbelief}
\end{equation}
It is unchanged by further zeros. The probabilities of output one at
these two contexts are $s_\varepsilon$ and $p$, respectively. Such an
output leads to $q_3^{\mathrm f}$ or $q_3^{\mathrm r}$, where the
belief on $(F_d,F_{d-1})$ has first coordinate $c_\varepsilon$ or $c_0$.
Output one from $q_0$ leads to the known conformation $F_d$ at $q_2$.
The remaining outputs resolve these mixtures and complete the normal
countdown. Table~\ref{tab:plateaufilter} records the resulting filter.

\begin{table}[t]
\caption{Output filter for \eqref{eq:plateaupolicy} after the first
return to $q_0$. The probability column gives the probability of output one conditional on
the context listed in the first column. A dash denotes an impossible output, and $u_0=q_0$.}
\label{tab:plateaufilter}
\centering
\small
\setlength{\tabcolsep}{5pt}
\begin{tabular}{@{}lccc@{}}
\toprule
Context & $P(D_t=1)$ & After $0$ & After $1$\\
\midrule
$q_0$ & $k_\varepsilon$ & $q_1^{\mathrm f}$ & $q_2$\\
$q_1^{\mathrm f}$ & $s_\varepsilon$ & $q_1^{\mathrm r}$ & $q_3^{\mathrm f}$\\
$q_1^{\mathrm r}$ & $p$ & $q_1^{\mathrm r}$ & $q_3^{\mathrm r}$\\
$q_2$ & $1$ & --- & $u_{d-1}$\\
$q_3^{\mathrm f}$ & $c_\varepsilon$ & $u_{d-2}$ & $u_{d-1}$\\
$q_3^{\mathrm r}$ & $c_0$ & $u_{d-2}$ & $u_{d-1}$\\
$u_k$, $1\le k<d$ & $0$ & $u_{k-1}$ & ---\\
\bottomrule
\end{tabular}
\end{table}

These updates also cover $d=2$: the zero branch from either $q_3$
context is then a use of $F_1$ and returns directly to the mixture in
\eqref{eq:plateauresetbelief}. The split contexts retain the same
original $Q$-node, so their projection follows \eqref{eq:Qgraph}.
At $\varepsilon=0$, their predictive probabilities reduce to
$k_0=p/2$, $s_0=p$, and $c_0=(\lambda+1)^{-1}$.

For all sufficiently small $\varepsilon\ge0$, every displayed
nondeterministic probability is strictly between zero and one. The
finite context chain is irreducible and has a self-loop at
$q_1^{\mathrm r}$. Let $\mu_\varepsilon$ denote its stationary law.
Conditional entropy and divergence rates therefore exist from the
specified initial state; the initial finite-mean transient has no effect
on either rate.

\subsection{Information-rate expansion}

Relative to the original test process, only the three predictive rows
at $q_0,q_1^{\mathrm f},q_3^{\mathrm f}$ change. The relative-entropy
chain rule gives
\begin{align}
 \mathcal D_a(\varepsilon)
 &:={\lim_{n\to\infty}}\frac1nD(P_{D^n}\|T_{D^n})\nonumber\\
 &=\mu_\varepsilon(q_0)D_{\rm B}(k_\varepsilon\|p/2)\nonumber\\
 &\quad+\mu_\varepsilon(q_1^{\mathrm f})
                       D_{\rm B}(s_\varepsilon\|p)\nonumber\\
 &\quad+\mu_\varepsilon(q_3^{\mathrm f})
                       D_{\rm B}(c_\varepsilon\|c_0)
 =O(\varepsilon^2).
 \label{eq:plateauklrate}
\end{align}
To justify the last equality, each first argument is a smooth function
of $\varepsilon$ equal to its fixed second argument at zero. For
$b\in(0,1)$,
\begin{equation}
 D_{\rm B}(b+t\|b)
 =\frac{t^2}{2\ln 2\,b(1-b)}+O(t^3).
 \label{eq:plateauklextension}
\end{equation}
The three stationary weights lie in $[0,1]$, so no differentiation of
those weights is needed for \eqref{eq:plateauklrate}.

Let $\pi_\varepsilon$ be the stationary conformation law under
\eqref{eq:plateaupolicy}, and write
$w_\varepsilon=\pi_\varepsilon(F_1)$. The deterministic normal countdown gives the
same mass $w_\varepsilon$ to every $F_k$. Flow balance at $J$ and $R$
gives
\begin{align}
 \pi_\varepsilon(J)&=\varepsilon w_\varepsilon,\nonumber\\
 p\pi_\varepsilon(R)&=(1-a\varepsilon)w_\varepsilon.
 \label{eq:plateauflows}
\end{align}
Normalization then yields
\begin{align}
 w_\varepsilon&=\frac{p}{1+dp+\varepsilon(p-a)},\nonumber\\
 \eta_\varepsilon&=\frac{\varepsilon p}
                         {1+dp+\varepsilon(p-a)},
 \label{eq:plateaudensity}
\end{align}
where $\eta_\varepsilon$ is the premature-request density.

The Bellman substitutions of Appendix~\ref{app:bellman} remain
identities for all $a\in[0,1]$, including where the premature-action
residual is negative. Divide \eqref{eq:exactgap}, with $\nu=0$, by $n$
and take its limit under \eqref{eq:plateaupolicy}. Because $v$ is bounded, its endpoint contribution vanishes after division by $n$. The finite filter and finite physical-state chain
give
\begin{align}
 \mathcal I_a(\varepsilon)
 &:={\lim_{n\to\infty}}\frac1n I(U^n\!\to D^n)\nonumber\\
 &=\rho+\{a-H_2(a)\}\eta_\varepsilon
                -\mathcal D_a(\varepsilon).
 \label{eq:plateaurateidentity}
\end{align}
Equations~\eqref{eq:plateauklrate} and \eqref{eq:plateaudensity} prove
\eqref{eq:plateauderivative}.

\subsection{Operational achievability by resetting}

A sequence of $2d$ zero requests returns every conformation to $R$
with certainty. Normal recovery takes at most $d$ uses. A shadow
conformation takes at most $d-1$ uses to reach $J$, one further use
through $J$, and at most $d$ normal recovery uses if the toggle occurs.
Any remaining zero requests keep the conformation at $R$.

For a fixed $n$, let a strategy tree $V$ specify one binary request for
each possible output history of length less than $n$. Generate its
nodes recursively according to \eqref{eq:plateaupolicy}, evaluating the
conformation from the input and output history along that branch.
Selecting $V$ before transmission and following it induces exactly the
desired randomized causal law. Given $V$ and the past outputs, all past
requests and the current conformation are fixed. The conditional chain
rule therefore gives
\begin{equation}
 I(V;D^n)=I(U^n\!\to D^n).
 \label{eq:plateautreeidentity}
\end{equation}

Regard $V$ as the input of a channel used for $n$ clock periods, followed
by the $2d$-use reset. Start each such block in $R$ and discard its reset
outputs at the decoder. The resulting channel from strategy trees to
length-$n$ output blocks is memoryless. Ordinary channel coding
\cite{Shannon1948} attains every rate below
\begin{equation}
 \frac{I(V;D^n)}{n+2d}
 =\frac{I(U^n\!\to D^n)}{n+2d}
 \label{eq:plateautreebound}
\end{equation}
per original channel use. The selected tree is fixed by the message in
each block, so this construction does not require private encoder
randomness in the resulting code. Taking $n\to\infty$ proves
$C_{\mathrm{fb}}\ge\mathcal I_a(\varepsilon)$ and completes the
achievability step in Theorem~\ref{thm:capacityplateau}.

\section{Resource-constrained realizations}
\label{app:weighted}

The Lagrangian form of dynamic matching admits a constructive resource extension. We give the coding argument because a stationary entropy calculation alone does not establish zero-error capacity at a fixed cost budget.

\subsection{Weighted paths and their operational rates}

Let $G=(V,E)$ be finite, irreducible, and right resolving, with fixed initial vertex $i_0$ and edge costs $c(e)\ge0$. For $\theta\ge0$ define
\begin{equation}
 A_\theta(i,j)=\sum_{e:i\to j}2^{-\theta c(e)},\quad
 A_\theta r_\theta=\Lambda_\theta r_\theta,
 \label{eq:app-weightedPF}
\end{equation}
where $r_\theta>0$. Set $P(\theta)=\log_2\Lambda_\theta$ and $h_\theta(i)=\log_2(r_\theta)_i$. The tilted edge law is
\begin{equation}
 \pi_\theta(e\mid i)=
 \frac{2^{-\theta c(e)}(r_\theta)_{\head(e)}}{\Lambda_\theta(r_\theta)_i}.
 \label{eq:app-tilted}
\end{equation}
Let $\Pi_{\theta,i_0}^{(n)}$ be its law on length-$n$ paths starting from $i_0$. Let $P_\theta$ be its vertex transition matrix, $\mu_\theta$ its stationary edge flow, and
\begin{equation}
 B_\theta=\sum_e\mu_\theta(e)c(e).
\end{equation}
Since every edge has positive tilted probability, the stationary vertex law exists uniquely and is positive, without an aperiodicity assumption.

Taking logarithms of \eqref{eq:app-tilted} and summing along a path gives the exact identity
\begin{align}
 H(E^n)-\theta\mathbb E\sum_{t=1}^n c(E_t)
 &+D(P_{E^n}\|\Pi_{\theta,i_0}^{(n)})\nonumber\\
 &=nP(\theta)+h_\theta(i_0)-\mathbb E h_\theta(V_n).
 \label{eq:app-weighted-path}
\end{align}
In particular, the tilted entropy rate is
\begin{equation}
 \mathcal H(\mu_\theta)=P(\theta)+\theta B_\theta.
 \label{eq:app-weightedrate}
\end{equation}
A stationary edge flow is a probability distribution $\mu$ on $E$ with equal incoming and outgoing mass at every vertex:
\begin{equation*}
 \mu_i=\sum_{e:\tail(e)=i}\mu(e)=\sum_{e:\head(e)=i}\mu(e).
\end{equation*}
Its entropy rate is
\begin{equation}
 \mathcal H(\mu)=-\sum_e\mu(e)\log_2\frac{\mu(e)}{\mu_{\tail(e)}}.
 \label{eq:flowentropy}
\end{equation}
The convention for zero terms makes this a continuous function of the stationary flow.

\begin{lemma}[Fixed-budget path codes]
\label{lem:costcodes}
Let $\mu$ be a positive stationary edge flow on $G$, with mean cost $B$. Suppose a stationary edge flow of cost strictly below $B$ exists. Then for each $\varepsilon>0$ and all sufficiently large blocklengths there is a set of distinct legal label strings from $i_0$, each of average cost at most $B$, whose rate is at least $\mathcal H(\mu)-\varepsilon$.
\end{lemma}

\begin{IEEEproof}
Let $\mu^-$ be a lower-cost stationary flow. For sufficiently small $\eta>0$, the balanced flow
$\mu^{(\eta)}=(1-\eta)\mu+\eta\mu^-$ is positive, has cost strictly below $B$, and has entropy arbitrarily close to $\mathcal H(\mu)$. The stationary-flow constraints have integer coefficients. Rational points are therefore dense in their polytope, so a positive rational balanced flow $q$ can be chosen with cost $B_q<B$ and $\mathcal H(q)>\mathcal H(\mu)-\varepsilon/2$. This also follows by decomposing flows into directed cycles and approximating their mixing weights by rationals.

Choose a denominator $N$ for $q$. For any multiple $m$ of $N$, let $n_e=mq(e)$ and $n_i=\sum_{e:\tail(e)=i}n_e$. These are positive integer balanced counts. Consider edge strings of length $m$ beginning and ending at $i_0$ with these counts. Their number is
\begin{equation}
 |\mathcal T(n_e)|=\tau_{i_0}(n_e)
 \frac{\prod_i(n_i-1+\mathbf1\{i=i_0\})!}{\prod_e n_e!},
 \label{eq:Euler-type}
\end{equation}
where $\tau_{i_0}(n_e)$ is the number of directed spanning trees toward $i_0$ in the multigraph with $n_e$ distinguishable copies of edge $e$.

For completeness, order the outgoing edge copies at each vertex according to their occurrence in a tour. The final exit from every vertex except $i_0$ must form a directed tree toward $i_0$. Conversely, any such tree and orders of the remaining copies generate an Euler tour by following the next unused copy; a premature stop is excluded by the last-exit tree. There are $n_{i_0}!$ orders at the root and $(n_i-1)!$ at each other vertex. Summing over trees counts tours with distinguishable copies. Each edge-type string has exactly $\prod_e n_e!$ assignments of copies, proving \eqref{eq:Euler-type}.

The tree factor grows polynomially in $m$. Stirling's formula therefore gives
\begin{equation}
 \log_2|\mathcal T(n_e)|=m\mathcal H(q)+O(\log m).
 \label{eq:typeexponent}
\end{equation}
Every path in this set has total cost $mB_q$ and terminal vertex $i_0$. Because the graph is right resolving and the initial vertex is fixed, distinct paths give distinct label strings.

For an arbitrary large length $n$, take $m=N\lfloor n/N\rfloor$ and append one fixed legal path of length $n-m$ from $i_0$. Its additional cost is at most $N\max_ec(e)$, independent of $n$. The strict margin $B-B_q$ then makes every completed string satisfy the budget for all sufficiently large $n$. Appending the same suffix preserves distinctness, and its bounded length does not change the asymptotic rate. Equation~\eqref{eq:typeexponent} proves the claim.
\end{IEEEproof}

Let $[B_{\min},B_{\max}]$ be the interval of costs of stationary edge flows. At a tilted point with $B_\theta\in(B_{\min},B_{\max})$, Lemma~\ref{lem:costcodes} applies to $\mu_\theta$. Equation~\eqref{eq:app-weighted-path} gives the matching converse. The zero-error rate of the noiseless constrained system at that point is therefore $P(\theta)+\theta B_\theta$, under either a per-codeword or expected-average budget. This argument handles periodic graphs and fixed starting vertices directly.

\subsection{Soft channels at a prescribed cost boundary point}

Use the universal architecture of Section~\ref{sec:constraints}, replacing $\pi^*$ by $\pi_\theta$ in the initiating and autonomous rows. Give a desired controlled edge $e$ cost $c(e)$; give an initiating undesired input and every autonomous use a constant physical cost $\beta\ge0$. The cost is a function of the channel state and input, not the receiver context.

\begin{theorem}[Resource-matched soft realization]
\label{thm:weightedsoft}
Suppose the Perron root $\lambda$ of the unweighted adjacency matrix $A$ exceeds one and $B_\theta$ is in the interior of the stationary-cost interval. Choose an integer $L$ such that every vertex has at least two legal length-$L$ paths and
\begin{equation}
 L\{P(\theta)+\theta\beta\}>\Span(h_\theta).
 \label{eq:weightedL}
\end{equation}
Then the constructed channel satisfies
\begin{equation}
 C_{0,\mathrm{fb}}(B_\theta)=C_{\mathrm{fb}}(B_\theta)
 =P(\theta)+\theta B_\theta.
 \label{eq:weighted-cap}
\end{equation}
Every undesired controlled action has Lagrangian residual at least
$L\{P(\theta)+\theta\beta\}-\Span(h_\theta)>0$.
\end{theorem}

\begin{IEEEproof}
Use the test edge law $\pi_\theta$, the multiplier $\nu=\theta$, the average Lagrangian reward $\rho=P(\theta)$, and potential
\begin{equation}
 v_\theta(i,m)=P_\theta^m h_\theta(i)-m\{P(\theta)+\theta\beta\}.
 \label{eq:weightedpotential}
\end{equation}
For a desired deterministic edge $e:i\to j$,
\begin{equation}
 -\log_2\pi_\theta(e\mid i)-\theta c(e)+h_\theta(j)
 =P(\theta)+h_\theta(i).
\end{equation}
For an autonomous row, the divergence reward is zero and the physical cost is $\beta$. The recursion in \eqref{eq:weightedpotential} gives
\begin{equation}
 -\theta\beta+\mathbb E v_\theta(V^+,m-1)
 =P(\theta)+v_\theta(i,m).
\end{equation}
For an initiating undesired row the same calculation gives residual
\begin{equation}
 \Delta_{\theta,L}(i)
 =L\{P(\theta)+\theta\beta\}
    +h_\theta(i)-P_\theta^Lh_\theta(i).
 \label{eq:weightedgap}
\end{equation}
It is positive by \eqref{eq:weightedL}. Theorem~\ref{thm:exactgap} gives the cost-constrained upper bound in \eqref{eq:weighted-cap}. Lemma~\ref{lem:costcodes} applied to the legal subchannel gives zero-error codes at that rate, with every codeword satisfying the same budget. This proves equality. Positive tilted probabilities on at least two length-$L$ paths make the initiating action behaviorally distinct, by the same common-continuation argument used for Theorem~\ref{thm:universal}.
\end{IEEEproof}

For $\theta>0$, choose $\beta$ so that $P(\theta)+\theta\beta>0$, then take $L$ large enough to meet both conditions. Irreducibility and $\lambda>1$ ensure multiple paths from every vertex at all sufficiently large lengths. At $\theta=0$, the condition reduces to the earlier unweighted one. If every edge has cost $c_0$, the unweighted legal codes already meet budget $c_0$ exactly, so Lemma~\ref{lem:costcodes} is unnecessary. Changing $\theta$ generally changes the channel; the theorem realizes boundary points across a family and does not determine the full capacity--cost curve of one fixed soft channel.

\end{document}